\documentclass[11pt]{article}
\usepackage{amssymb, amsfonts, graphicx, bm, amsmath, amsthm}
\usepackage{color, multirow, booktabs}
\usepackage[]{natbib}

\newtheoremstyle{sjsstyle1}{\medskipamount}{\medskipamount}{\normalfont}{}{\bfseries}{.}{1em}{}
\newtheoremstyle{sjsstyle2}{\medskipamount}{\medskipamount}{\normalfont}{}{\itshape}{.}{1em}{}

\theoremstyle{plain}
\renewcommand\theequation{\thesection.\arabic{equation}}

\newtheorem{theorem}{Theorem}[section]
\newtheorem{lemma}[theorem]{Lemma}
\newtheorem{proposition}[theorem]{Proposition}

\theoremstyle{sjsstyle1}
\newtheorem{definition}[theorem]{Definition}

\theoremstyle{sjsstyle2}
\newtheorem{remark}[theorem]{Remark}
\newtheorem{example}{Example}
\makeatletter \@addtoreset{equation}{section} \makeatother

\newcommand{\pr}{\mbox{pr}}
\newcommand{\cov}{\mbox{cov}}
\newcommand{\var}{\mbox{var}}
\newcommand{\mc}[1]{\mathcal {#1}}
\newcommand{\bb}[1]{\mathbb {#1}}
\newcommand{\eps}{\varepsilon}

\newcommand{\ZZ}{\mathbb Z}
\newcommand{\RR}{\mathbb R}

\newcommand{\rcs}[1]{\bm{#1}}
\newcommand{\wat}{\widehat}
\newcommand{\oar}{\overline}

\begin{document}

\title{Confidence Regions for Means of Random Sets using Oriented Distance Functions}
\author{Hanna K. Jankowski \\
York University\\
        \and
        Larissa I. Stanberry\\
        University of Washington\\\\}
\date{February 25, 2011}

\maketitle
\begin{abstract}
Image analysis frequently deals with shape estimation and image reconstruction.  The objects of interest in these problems may be thought of as random sets, and one is interested in finding a representative, or expected, set.  
We consider a definition of set expectation using oriented distance functions and study the properties of the associated empirical set.  Conditions are given such that the empirical average is consistent, and a method to calculate a confidence region for the expected set is introduced.   The proposed method is applied to both real and simulated data examples.\\

\end{abstract}

\section{Introduction}

Image analysis frequently deals with the problem of shape estimation.  Many instances of this may be found, for example, in medical imaging, where shape analysis of brain structures is often used to differentiate between different populations of subjects.  Examples include the study of the hippocampus for schizophrenic patients and corpus callosum for adults with fetal alcohol exposure, as well as other neuroanatomical structures \citep{styner:04, bookstein:02, bookstein:02b, styner:03, levitt:09}.

To analyze the observed shapes, it is natural to think of them as realizations of random sets.  The problem of shape inference then translates into finding the average set and describing its variability.  

As the space of closed sets is nonlinear, there is no natural way to define the mean of a set.   Indeed, many different definitions of the expected set exist, and therefore one must first select a definition of the mean to work with.  Quoting \cite{molchanov:05}, ``the definition of the expectation depends on what the objective is, which features of random sets are important to average, and which are possible to neglect".  Here, we focus on the definition of set expectation given in \cite{stanberry:j:08}, which is based on the oriented distance function (ODF) \footnote{To differentiate it from others, we will refer to the \cite{stanberry:j:08} definition as the ODF definition.  However, the definition \cite{baddeley:molchanov:98} may also be based on the ODF.}.    The definition is akin to those considered in \cite{baddeley:molchanov:98, lewis:99}, which also rely on the distance function.  The ODF definition, however, has several desirable theoretical properties and was found to outperform other definitions in image applications; for a detailed comparison see \citet{stanberry:j:08}.  A thorough review of other existing definitions appears in \cite{molchanov:05}.  


In this paper, we study the empirical estimators of the expected set and expected boundary using the ODF definition.  We give conditions for consistency of these estimators.  We also study their variability using the concept of confidence regions.

\cite{seri:c:04} present two methods for the calculation of the confidence region of the Aumann expectation.   To our best knowledge, this is the only other time confidence regions were studied in the context of random sets.  The methodology was developed for convex sets, and therefore the Aumann expectation is a natural choice, as it always yields a convex answer.


\citet[Theorem 3.1]{molch:98} gives a central limit theorem for the Hausdorff distance of sublevel sets.  \citet{baddeley:molchanov:98} use this result to obtain a central limit theorem for their expectation, and this may also be done for the ODF definition.   Such central limit theorems could potentially be used to calculate confidence regions.   The regions may be found as a dilation of the empirical set estimator.   However, this approach requires the estimation of a complex functional of the derivative of the expected distance transform, which renders the method impractical.   Furthermore, a dilation approach would provide a uniform confidence region:  the distance between the boundary of the mean set and the boundary of the confidence set would be equivalent at all points.  Thus, the dilation method would mask important information about the local variability of the estimators.

In this paper, we propose a new and simple approach to calculate confidence regions for both the mean set and its boundary.  The method works for both convex and non-convex sets.    The resulting confidence regions are conservative in that they cover the expected set with {at least} $100(1-\alpha)\%$ probability.  We show that the confidence regions satisfy certain natural equivariance properties, which are analogous to those of confidence intervals for real--valued parameters.   Moreover, the confidence regions provide a simple visual representation of the variability of the estimators and are able to detect local changes in variability.  The method can also be used in Bayesian inference to study the behaviour of the posterior sample.

Our definition of the confidence region is based on the quantiles of the supremum of a Gaussian random field. We consider several examples where this quantile is calculated easily.  When these quantiles are not available analytically, we propose a bootstrap method to provide approximate confidence regions.  The bootstrap approach also avoids making an assumption of an underlying distribution of the observed sets,  which could be quite difficult for practitioners.  


The outline of this paper is as follows.  In Section \ref{sec:set_mean}, we review the definitions of set and boundary expectations given in \cite{stanberry:j:08}.  Consistency is studied in Section \ref{sec:consistent} and the confidence regions are described in Section~\ref{sec:confidence_odf}. 
 Section~\ref{sec:examples} gives several examples of our approach, including a simulation study.  We consider the toy image reconstruction example discussed by \cite{baddeley:molchanov:98}, and analyse the sand grains data first discussed in \cite{stoyan:97}.    
 Lastly, we consider a medical imaging example, and apply our methods to a boundary reconstruction problem in a mammogram image.    Proofs and technical details appear in the Appendix, which is available online as supplementary material.

\subsection{Notation and Assumptions}

Throughout, we let $\mc D$ denote the domain on which the sets are observed.  Unless otherwise stated, we assume that $\mc D$ is the working domain and write, for example, $A=\{x: x \in A\}$ without stating that $x\in \mc D$ explicitly.  We also assume that $\mc D$ is a subset of $\RR^d$, and denote the Euclidean norm of $x$ as $|x|$.

We write $B_r(x_0)=\{x: |x-x_0|\leq r\}$ for the closed ball of radius $r$ centered at $x_0$.    For a set $A$, we write $A^o, \overline{A}, A^c$ and $\partial A$ to denote its interior, closure, complement and boundary.   Unless noted otherwise, set operations are calculated relative to the domain $\mc D$.  That is, $A^c = A^c \cap \mc D,$ and so forth.  
Deterministic sets are denoted using capital letters  $A, B \ldots$, while bold upper-case lettering, $\rcs A, \rcs B, \ldots$, is used for random sets.

The notation $C(\mc D)$ is used to denote the space of continuous functions $C(\mc D)=\{ f:\mc D\mapsto\RR, \ f \mbox{ continuous}\}$ endowed with the uniform topology.  That is, $f_n\rightarrow f$ in $C(\mc D)$ if $\sup_{x\in K}|f_n(x)-f(x)|\rightarrow 0$, for all compact subsets $K\subset \mc D.$  We write $X_n\Rightarrow X$ to say that $X_n$ converges weakly to $X$.  Throughout the paper, when handling weak convergence of stochastic processes or random fields, we assume that these take values in $C(\mc D)$.

\section{Random Closed Set and Its Expectation}\label{sec:set_mean}

Let $\mc F$ be the family of closed sets of $\RR^d$ and let $\mc K$ denote the family of all compact subsets of $\RR^d$.   
For a probability triple $(\Omega, \mc A, P)$, a random closed set (RCS) is the mapping $\rcs A:\Omega\mapsto \mc F$ such that for every compact set $K\in \mc K$
\begin{eqnarray*}
\{\omega: \rcs A(\omega)\cap K \neq \emptyset\}\in \mc A.
\end{eqnarray*}
For more background on random closed sets, we refer to \cite{matheron:75, ayala:91}.

\subsection{Definition of Expectation}

For a nonempty set $A\subset \mc D,$ the distance function is defined  as $d_A(x)= \inf_{y\in A} |x-y|$ for all $x\in\mc D.$  
Given the distance function of a closed set $A$, the original set may be recovered via $A=\{x: d_A(x)=0\}$.  Also, the Hausdorff distance may be calculated using the distance function as
\begin{eqnarray*}
\rho(A, B) &=& \max\left\{\sup_{x\in A }d(x, B), \sup_{x\in B}d(x, A)\right\}\\
      &=& \sup_{x\in \mc D}|d_A(x)-d_B(x)|,
\end{eqnarray*}
for any sets $A, B \subset \mc D$.  
We refer to \cite{delfour:zolesio:02} for more mathematical properties of the distance function.  The distance function has also been used in the context of image thresholding, see for example \cite{friel:molchanov:99, molchanov:teran:03}.

The oriented distance function (ODF) of any set $A\subset \mc D$ such that $\partial A\neq 0$ is defined as $b_A(x)= d_A(x)-d_{A^c}(x)$ for all $x\in\mc D.$   
For a closed set, the set and its boundary may now be recovered 
by $A=\{x: b_A(x)\leq 0\}$ and $\partial A=\{x: b_A(x)=0\}$.   Note that, given {only} the information at a fixed point $x_0,$ the ODF is more informative than the distance function.   Given the value of $b_A(x_0)$ we know the value of $d_A(x_0),$ but the converse statement is not true.

For an RCS $\rcs A$ with a.s. non-empty boundary we define the random function $b_{\rcs A}(x)$, and denote its pointwise mean as $E[b_{\rcs A}(x)]$.  The mean set and mean boundary are then defined as follows.
\begin{definition}\label{def:meanset}
Suppose that $\rcs A$ is a random closed set such that $\partial \rcs A\neq \emptyset$ almost surely and assume that $E[|b_{\rcs A}(x_0)|]<~\infty$ for some $x_0\in \mc D$.  Then
\begin{eqnarray*}
  E[\rcs A] &=& \{x: E[b_{\rcs A}(x)]\leq 0\}, \hspace{0.5cm}\\
  \Gamma[\rcs A] &=& \{x: E[b_{\rcs A}(x)]=0\}.
\end{eqnarray*}
Furthermore, we define the expectation of the complement as $E[\rcs A^c]=\{x: E[b_{\rcs A}(x)]\geq 0\}.$
\end{definition}

\begin{remark}
If $A$ is closed and $\partial A\neq\emptyset$ then $A^c$ is open, however, the oriented distance function of $A^c$ continues to be well--defined, and indeed we have that $b_{A^c}(x)=~-~b_{A}(x)$.  Thus, $E[\rcs A^c]=E\left[\overline{\rcs A^c}\right]=\{x: E[b_{{\rcs A}}(x)]\geq 0\}$. 
\end{remark}

\begin{example}[disc with random radius]\label{ex:randomcircle}
Suppose that $\rcs A=\{x: |x|\leq R\} \subset \RR^d$, for some non--negative, integrable, real-valued random variable $R$.  Then $b_{\rcs A}(x)=|x|-R$ and hence $E[\rcs A]=\{x: |x|\leq E[R]\}$ and $\Gamma[\rcs A] = \{x: |x|=E[R]\}$.  That is, the expected set is a disc with radius $E[R]$, with boundary equal to the expected boundary.
\end{example}

\begin{example}[random singleton]\label{ex:randomsingleton}
Suppose that $\rcs A=\{X\}$ for some $\RR^d$-valued random variable $X$, then $E[\rcs A]=\Gamma[\rcs A]=\emptyset.$
\end{example}

In Example \ref{ex:randomcircle}, Definition \ref{def:meanset} yields a natural answer, while the result of Example \ref{ex:randomsingleton} seems counterintuitive.  As mentioned previously, the choice of expectation depends on the problem at hand.  The ODF definition was motivated by problems in imaging, where random singletons are regarded as noise.  Example \ref{ex:randomsingleton} illustrates a natural de-noising property of the expectation; see also the discussion in \cite{stanberry:j:08}.  

The function $E[b_{\rcs A}(x)]$ provides additional information about the mean and its boundary.  Recall that $b_{\rcs A}(x)$ is Lipschitz with constant one almost surely \citep{delfour:zolesio:96}.  Therefore, $E[b_{\rcs A}]$ is also Lipschitz, which also implies that $E[\rcs A]$ and $\Gamma[\rcs A]$ are closed sets.  Furthermore, if the function $E[b_{\rcs A}(x)]$ is smooth in a neighbourhood of $x_0$, then $\Gamma[\rcs A]$ is also smooth near $x_0.$

\begin{proposition}\label{prop:smoothboundary}
Suppose that $d\geq 2,$ and fix $k\geq 1$, and suppose that the function $E[b_{\rcs A}(x)]$ is $C^k$ in a neighbourhood of $x_0$, and that $|\nabla E[b_{\rcs A}(x_0)]|\neq 0.$  Then, there exists a (possibly different) neighbourhood of $x_0$ such that $\Gamma[\rcs A]$ is $C^k$.
\end{proposition}

We next consider the estimation of $E[\rcs A]$ and $\Gamma[\rcs A].$

\begin{definition}
Suppose that we observe  $\rcs A_1, \ldots, \rcs A_n$ random sets, and let
\begin{eqnarray*}
& \bar b_n(x)= \frac{1}{n}\sum_{i=1}^n b_{\rcs A_i}(x). &
\end{eqnarray*}
Then the empirical mean set and the empirical mean boundary are defined as
\begin{eqnarray*}
  \bar {\rcs A}_n = \{x: \bar b_n(x) \leq 0\} \ \mbox{ and } \
  \oar{\rcs \Gamma}_n = \{x: \bar b_n(x) = 0\}.
\end{eqnarray*}
\end{definition}

Similarly to the mean ODF, the function $\bar b_n$ is Lipschitz, and therefore the empirical sets $\bar {\rcs A}$ and $\oar{\rcs \Gamma}_n$ are both closed.

\subsection{Consistency and Fluctuations of the Empirical ODF}

Suppose that we have $n$ independent and identically distributed (IID) RCSs $\rcs A_1, \ldots, \rcs A_n$.  Then the following results are valid for the average ODF.

\begin{theorem}\label{thm:LLN}
Suppose that $\rcs A_1, \ldots, \rcs A_n$ are IID, and that for some $x_0\in \mc D$, $E[b_{\rcs A}(x_0)]<\infty.$  Then  for all compact subsets $K\subset\mc D$
\begin{eqnarray*}
\lim_n\sup_{x\in K}|\bar b_n(x)-E[b_{\rcs A}(x)]|=0,
\end{eqnarray*}
almost surely.
\end{theorem}

\begin{theorem}\label{thm:randomsetCLT}
Suppose that $\rcs A_1, \ldots, \rcs A_n$ are IID, and that $E[b_{\rcs A}^2(x_0)]<\infty$ for some $x_0\in \mc D$.  Then
\begin{eqnarray*}
\ZZ_n(x) \equiv  \sqrt{n}(\bar b_n(x)-E[b_{\rcs A}(x)]) \Rightarrow \ZZ(x),
\end{eqnarray*}
where $\ZZ$ is a mean zero Gaussian random field with covariance
\begin{eqnarray*}
cov(\ZZ(x), \ZZ(y))= E[b_{\rcs A}(x)b_{\rcs A}(y)]-E[b_{\rcs A}(x)]E[b_{\rcs A}(y)]
\end{eqnarray*}
for $x,y\in \mc D.$
\end{theorem}

The next result shows that $\ZZ$ has very smooth sample paths.  Recall that a function $f: \mathbb R^d \mapsto \mathbb R$ is H\"older of order $\alpha$ if it satisfies
$
|f(x)-f(y)| \leq K |x-y|^\alpha,
$
for some positive finite constant $K$ and $\alpha>0$, for all $x,y$ in the domain of~$f$.
\begin{proposition}\label{prop:lip}
For any $x,y, x', y' \in \mc D$
\begin{eqnarray}\label{line:VarBounds}
\var(\ZZ(x)-\ZZ(y))&\leq& |x-y|^2\label{line:VarBounds},\\
|\cov(\ZZ(y)-\ZZ(x),\ZZ(y')-\ZZ(x'))|&\leq & 2|y-x||y'-x'|\label{line:coVarBounds}.
\end{eqnarray}
Moreover, the sample paths of $\ZZ$ are H\"older of order $\alpha$, for any $\alpha<1$.
\end{proposition}


\section{Consistency}\label{sec:consistent}

Here, we study consistency of the estimators $\bar{\rcs A}_n$ and $\overline{\rcs \Gamma}_n$, assuming that $\bar b_n \rightarrow E[b_{\rcs A}]$ almost surely in $C(\mc D).$  By Theorem \ref{thm:LLN}, these results apply to IID random closed sets.  Following \linebreak \cite{molch:98}, we say that ${\rcs A}_n$ converges strongly to $A$, if the Hausdorff distance \linebreak $\rho(\rcs A_n\cap K, A\cap K)\rightarrow 0$ almost surely, for any compact set $K$. 
The following theorem follows from  \citet[Theorem 2.1]{molch:98} and \citet[Theorem 1]{cuevas:g:w:r:06}.

\begin{theorem}\label{thm:consistent}
Suppose that $$\lim_n\sup_{x\in \mc D}|\bar b_n(x)-E[b_{\rcs A}(x)]|=0$$ almost surely, and that $E[\rcs A]$ is well-defined.  Suppose also that the expected ODF, $E[b_{\rcs A}(x)]$, satisfies
\begin{eqnarray}\label{cond:consistent_mean_A}
\{x: E[b_{\rcs A}(x)] \leq 0\} &=& \overline{\{x: E[b_{\rcs A}(x)]<0\}}.
\end{eqnarray}
Then $\bar {\rcs A}_m$ converges strongly to $E[{\rcs A}]$.
If $E[b_{\rcs A}(x)]$ also satisfies
\begin{eqnarray}
\{x: E[b_{\rcs A}(x)] \geq 0\} &=& \overline{\{x: E[b_{\rcs A}(x)]>0\}},
\label{cond:consistent_mean_B}
\end{eqnarray}
then $\overline{\rcs\Gamma}_n$ converges strongly to $\Gamma[{\rcs A}]$.
\end{theorem}

Condition \eqref{cond:consistent_mean_A} says that the expected ODF is not allowed to have a local minimum on $\Gamma[\rcs A]=\{x: E[b_{\rcs A}(x)]=0\}$, while \eqref{cond:consistent_mean_B} excludes mean ODFs which have a local maximum on $\Gamma[\rcs A]$ (again, this need not be unique).   Alternatively, since $E[b_{\rcs A}(x)]$ is a continuous, condition \eqref{cond:consistent_mean_A} says that  $E[\rcs A]$ is a topologically regular closed set, while condition \eqref{cond:consistent_mean_B} says that $E[\rcs A^c]$ is a topologically regular open set.

\begin{remark}
It is possible that $E[b_{\rcs A}(x)]$ violates the conditions \eqref{cond:consistent_mean_A} and/or \eqref{cond:consistent_mean_B} and consistency still holds.  For example, consider the RCS ${\rcs A}=\{x_0\}\subset \RR^d$ almost surely.  Then IID sampling trivially produces a consistent estimate, but $E[b_{\rcs A}(x)]$ fails to satisfy \eqref{cond:consistent_mean_A}.
\end{remark}

\begin{example}[half plane] For $\mc D\subset \RR^d$, consider the RCS ${\rcs A}=\{x\in \mc D: x_1 \leq \Theta\}$, where $\Theta$ is a real-valued random variable with finite mean $E[\Theta].$  Then $b_{\rcs A}(x)=x_1-\Theta$, and $E[b_{\rcs A}(x)]=x_1-E[\Theta].$  The mean ODF satisfies both conditions \eqref{cond:consistent_mean_A} and \eqref{cond:consistent_mean_B}, and therefore  $\bar {\rcs A}_n=\{x:x_1 \leq \bar \Theta_n\}$ and $\overline{\rcs\Gamma}_n=\{x: x_1 = \bar \Theta_n\}$ are consistent estimators of $E[{\rcs A}]=\{x: x_1 \leq E[\Theta]\}$ and $\Gamma[{\rcs A}]=\{x: x_1 = E[\Theta]\}$.  Indeed, we may easily check that $\rho(E[{\rcs A}], \bar {\rcs A}_n)=\rho(\Gamma[ {\rcs A}], \overline{\rcs\Gamma}_n)=|\bar \Theta_n-E[\Theta]|$ which converges to zero almost surely.
\end{example}

The following result provides some further insight into the consistency conditions.  

\begin{proposition}\label{prop:understanding_consistency}
Conditions \eqref{cond:consistent_mean_A} and \eqref{cond:consistent_mean_B} may be re--written in terms of mean set properties.
\begin{enumerate}
\item Condition \eqref{cond:consistent_mean_A} holds iff $\partial E[\rcs A^c]=\Gamma[\rcs A],$ and  iff $E[{\rcs A}] = \overline {(E[{\rcs A}^c])^c}.$
\item Condition \eqref{cond:consistent_mean_B} holds iff $\partial E[\rcs A]=\Gamma[\rcs A]$, and  iff $E[{\rcs A}^c] = \overline {(E[{\rcs A}])^c}.$
\end{enumerate}
\end{proposition}

\begin{example}[set and its boundary]\label{ex:setboundary}
Suppose that ${\rcs A}\subset \RR$ is either $[0,1]$ or $\{0,1\}$ with equal probability.  Then $E[{\rcs A}]=\Gamma[{\rcs A}]=[0,1]$, while $E[\rcs A^c]=\RR$.  On the other hand, if $[0,1]$ is seen with probability $p$, then if $p<0.5$, we have that $E[{\rcs A}]=\Gamma[{\rcs A}] = \{0,1\}$;  If $p>0.5$, then $E[{\rcs A}]=[0,1]$ and $\Gamma[{\rcs A}]=\{0,1\}.$  The case $p=0.5$ provides a setting where neither \eqref{cond:consistent_mean_A} nor \eqref{cond:consistent_mean_B} are satisfied.

We observe $n$ independent sets ${\rcs A}_1, {\rcs A}_2, \ldots, {\rcs A}_n$, where each $\rcs A_i$ is either $[0,1]$ or $\{0,1\}$ with equal probability.  Let $\wat p_n$ denote the proportion of the random sets  equal to $[0,1]$.  Then
$$
\bar b_n(x)=\wat p_n b_{[0,1]}(x)+(1-\wat p_n)b_{\{0,1\}}(x),
$$
and it follows that whenever $\wat p_n<0.5$,  $\bar {\rcs A}_n = \overline{\rcs\Gamma}_n=\{0,1\}$, and for $\wat p_n>0.5$,  $\bar {\rcs A}_n = [0,1]$ while $\overline{\rcs\Gamma}_n=\{0,1\}$.  Clearly, convergence to the expected set or the expected boundary can never be achieved.
\end{example}

\begin{example}[missing centre]\label{ex:timbit}
Suppose that $\rcs A$ is either a disc or an annulus in $\RR^2$; that is,
\begin{eqnarray*}
\rcs A&=& \left\{
             \begin{array}{ll}
             \{x: |x|\leq 1\} & \mbox{with probability  $p$},\\
             \{x: 0.5 \leq |x|\leq 1\} & \mbox{otherwise.}
             \end{array}
             \right.
\end{eqnarray*}
Then the expected set $E[{\rcs A}]$ is an annulus for $p< 1/3$, and a disc for $p\geq 1/3$.  For $p\neq 1/3$, the expected ODF satisfies both \eqref{cond:consistent_mean_A} and \eqref{cond:consistent_mean_B}.  For $p=1/3,$ we have
\begin{eqnarray*}
E[b_A(x)]&=& \left\{
             \begin{array}{ll}
             |x|-1 & \mbox{for } |x|\geq 0.75,\\
             -|x|/3 & \mbox{otherwise,}
             \end{array}
             \right.
\end{eqnarray*}
and hence $E[\rcs A]=\{x:|x|\leq 1\}$ while $\Gamma[\rcs A]=\{x:|x|=0,1\}\neq \partial E[\rcs A]$.  Since $E[b_{\rcs A}(x)]$ has a local maximum at $x=0$, it fails to satisfy \eqref{cond:consistent_mean_B}, and therefore this point may be omitted by the estimators~$\overline{\rcs \Gamma}_n$.
\end{example}

\begin{figure}[htb!]
\begin{center}
\fbox{\includegraphics[width=0.4\textwidth]{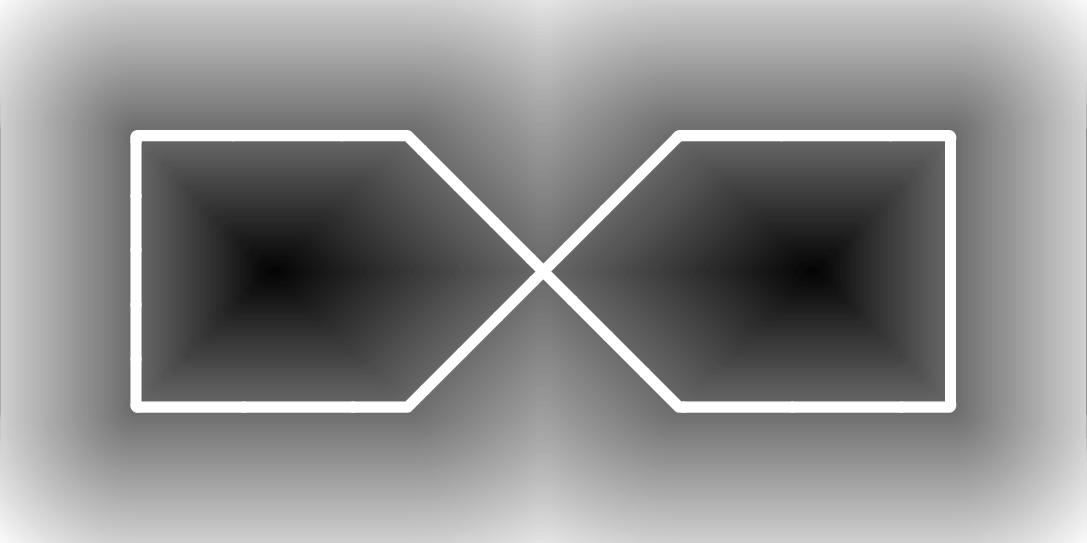}}
\end{center}
\caption{The expected boundary (white) for Example \ref{ex:blinkingsquare} is superimposed on a grey scale image of the expected ODF.}
\label{fig:blinkingsquare}
\end{figure}

\begin{example}[blinking square]\label{ex:blinkingsquare}
Suppose that the RCS ${\rcs A}$ is either a rectangle or a union of two squares with equal probability.  Specifically, define
\begin{eqnarray*}
A_1 &=& \{x: 0\leq x_1 \leq 3, 0\leq x_2 \leq 1\}, \\
A_2 &=& \{x: 0\leq x_1 \leq 1, 0\leq x_2 \leq 1\}
\cup\{x: 2\leq x_1 \leq 3, 0\leq x_2 \leq 1\}.
\end{eqnarray*}
Then $\rcs A = A_1$ with probability 0.5, and otherwise $\rcs A = A_2.$  Thus, half of the time the $\rcs A$ has its ``middle" removed.   The resulting mean set and mean boundary are shown in Figure~\ref{fig:blinkingsquare}.  Here, the expected ODF has no local maxima, or minima, along the boundary, and therefore both \eqref{cond:consistent_mean_A} and \eqref{cond:consistent_mean_B} are satisfied.
\end{example}

\section{Confidence Regions}\label{sec:confidence_odf}

We now construct confidence regions, or confidence supersets, for both $E[{\rcs A}]$ and $\Gamma[{\rcs A}].$  To do this, we assume that the sets are observed on a compact window $\mc W\subseteq \mc D$.  We also assume that $\ZZ_n\Rightarrow\ZZ$ in $C(\mc W),$ which holds, for example, under IID sampling by Theorem~\ref{thm:randomsetCLT}.

\begin{definition}
Let $q_1$ and $q_2$ be numbers such that \linebreak$\pr(\sup_{x\in \mc W} \ZZ(x) \leq q_1)=1-\alpha$, and
$\pr(\sup_{x\in \mc W} |\ZZ(x)| \leq q_2)=1-\alpha$.  Then, a $100(1-\alpha)\%$ confidence region for
$E[{\rcs A}]\cap \mc W$ is
\begin{eqnarray}\label{line:EACI}
\left\{x \in \mc W: \bar b_n(x) \leq q_1/\sqrt{n}\right\}
\end{eqnarray}
and a $100(1-\alpha)\%$ confidence region for
$\Gamma[{\rcs A}]\cap \mc W$ is
\begin{eqnarray}\label{line:EparACI}
\left\{x \in \mc W: |\bar b_n(x)| \leq q_2/{\sqrt{n}}\right\}.
\end{eqnarray}
\end{definition}

By  Proposition \ref{prop:lip}, the limiting field $\ZZ$ is continuous, and therefore both $\sup_{x\in \mc W} \ZZ(x)$ and $\sup_{x\in \mc W} |\ZZ(x)|$ are well-defined.  Further, Proposition \ref{prop:lip} gives a uniform upper bound on the variability of the increments of the random field.  Understanding the path properties of the process $\ZZ$, such as smoothness, provides information about the variability of the quantiles $q_1$ and $q_2$ and therefore also on the tightness of the confidence sets.  We also make the following comments about the new definition.

\begin{list}{}
        {\setlength{\topsep}{6pt}
        \setlength{\parskip}{0pt}
        \setlength{\partopsep}{0pt}
        \setlength{\parsep}{0pt}
        \setlength{\itemsep}{3pt}
        \setlength{\leftmargin}{20pt}}
  \item[1.]  The confidence region is conservative, in that it covers the set $E[{\rcs A}]\cap \mc W$ or $\Gamma[{\rcs A}]\cap\mc W$ with a probability of at least $100(1-\alpha)\%.$ One reason why the method is conservative is our use of the supremum of the fluctuation field to find the cut--off quantile values.  However, the field $\ZZ$ is very smooth and highly correlated by Proposition \ref{prop:lip}.  Therefore, we expect that the proposed method, although conservative, yields reasonable answers, especially for sets that have been co--registered apriori.  We explore the question of over--coverage via simulations in Section \ref{sec:sims}. 
  \item[2.]  The confidence region is ``immune" to the consistency conditions \eqref{cond:consistent_mean_A} and \eqref{cond:consistent_mean_B}; see Example \ref{ex:setboundary_CI}.
  \item[3.]  If the exact distribution of $\ZZ$ is unknown (as is often the case in practice), the quantiles may be approximated using a bootstrap approach.   For computational reasons, it is often easier to calculate asymmetric quantiles in \eqref{line:EparACI}.
  \item[4.]  The confidence region gives no information as to the geometry of the random sets.  That is, the shape of the confidence region may be similar for observed random sets which are stars, as well as for those which are squares.  
 \end{list}

\begin{center}
\begin{figure*}[htb!]
\includegraphics[width=1\textwidth]{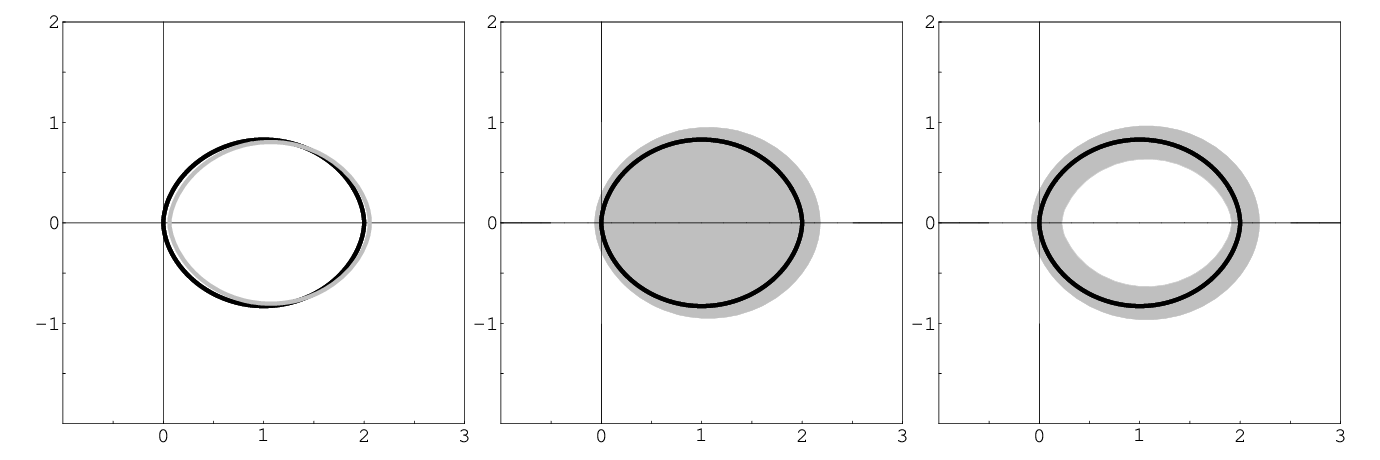}
\caption{Confidence regions for Example \ref{ex:circle_centre_CI}: (left) the mean boundary $\Gamma[{\rcs A}]$ in black and the empirical boundary $\overline{\rcs\Gamma}_n$ in grey; (centre) a 95\% bootstrap confidence set for $E[\rcs A]$;  (right) a 95\% bootstrap confidence set for $\Gamma[\rcs A]$.  The expected boundary $\Gamma[\rcs A]$ is shown in black for comparison.}
\label{fig:circleCI}
\end{figure*}
\end{center}

\begin{example}[disc in $\RR^2$ with random centre]\label{ex:circle_centre_CI}
The random set is a disc with radius one centred at $(\Theta,0)$ where $\Theta$ is Uniform$[0, 2]$, and suppose that we observe 100 IID random sets from this model.  The expected set $E[\rcs A]$ is shown in Figure~\ref{fig:circleCI}.  Moreover, since $E[b_{\rcs A}(x)]\geq |x-x_0|-1$ where $x_0=(E[\Theta],0)$, it follows that $E[\rcs A]$ is contained inside the disc of radius one centered at $(1,0)$. Confidence regions were formed for both $\Gamma[{\rcs A}]$ and $E[{\rcs A}]$ using re-sampling techniques to estimate the quantiles of $\sup_{x\in \mc W}\ZZ(x)$ and $\sup_{x\in \mc W}|\ZZ(x)|$ where the window $\mc W=[-2,2]\times[-1,3].$  These are illustrated in Figure \ref{fig:circleCI}.
\end{example}

\begin{example}[confidence regions for the set and its boundary]\label{ex:setboundary_CI} Let $[0,1] \subset \RR$ and suppose that $\rcs A$ is either $\{0,1\}$ or $[0,1]$ with equal probability.  Suppose also that we observe a simple random sample of size $n$ from this model.  Recall that $E[{\rcs A}]=\Gamma[{\rcs A}]=[0,1]$, and $E[b_{\rcs A}(x)]$ satisfies neither \eqref{cond:consistent_mean_A} nor \eqref{cond:consistent_mean_B}.  As before, let $\wat p_n$ denote the proportion of times that the set $[0,1]$ is observed.  If $\wat p_n>0.5$, then $\bar {\rcs A}_n = \overline{\rcs\Gamma}_n=\{0,1\}\neq [0,1]$.  If $\wat p_n < 0.5$, then $\bar {\rcs A}_n=[0,1]$ with $\overline{\rcs\Gamma}_n=\{0,1\}\neq \Gamma[{\rcs A}].$

The fluctuation field is given by
\begin{eqnarray*}
\ZZ_n(x)
        &=& \sqrt{n}(\wat p_n - 0.5)\left( b_{[0,1]}(x)- b_{\{0,1\}}(x)\right)\\
         &\Rightarrow&  Z \left( b_{[0,1]}(x)- b_{\{0,1\}}(x)\right),
\end{eqnarray*}
where $Z$ is a univariate normal random variable with mean zero and variance $0.25$.
The largest difference for $b_{[0,1]}(x)- b_{\{0,1\}}(x)$ occurs at $x=0.5$, and hence, for any window  such that $[0,1] \subset \mc W$, we have
\begin{eqnarray*}
\sup_{x\in \mc W} \ZZ(x) = \max\{-Z,0\}, \ \ \
\mbox{and } \ \sup_{x\in \mc W} |\ZZ(x)| = |Z|.
\end{eqnarray*}
Therefore, the exact quantiles are $q_1 = 1.645/2$ and $q_2=1.96/2$, and the confidence region for  $\Gamma[{\rcs A}]$ is given by $\{x: |\bar b_n(x)| \leq 1.96/2\sqrt{n}\}.$  
Now, for any $n$, $\max_{x\in[0,1]}|\bar b_n(x)|=|\bar b_n(0.5)|=|0.5-\wat p_n|$. Therefore the confidence region misses a part of $\Gamma[{\rcs A}]$ if and only if $|0.5-\wat p_n|>1.96/2\sqrt{n}$.  For large $n$, this happens with a probability of roughly $0.95$.  On the other hand, the Hausdorff distance $\rho(\Gamma[\rcs A], \overline{\rcs \Gamma}_n)=0.5$ whenever $\wat p_n \neq 0.5.$  This example illustrates that the confidence regions are not affected by the violation of the consistency conditions.

\end{example}

\begin{example}[confidence set for disc with random radius]
Suppose that $\rcs A$ is a disc with random radius $R$ with $\mu=E[R]$ and $\sigma^2=\var(R)$.  Then the expected set is a circle with radius $\mu$.  Also, the 95\% confidence interval for $E[{\rcs A}]$ is a circle with radius $\mu+1.645\sigma/\sqrt{n}$, while the 95\% confidence set for $\Gamma[{\rcs A}]$ is the band $\{x: \mu-1.96\sigma/\sqrt{n}\leq |x| \leq \mu+1.96\sigma/\sqrt{n}\}$.
\end{example}

\begin{figure*}[htb!]
\begin{center}
\includegraphics[width=1\textwidth]{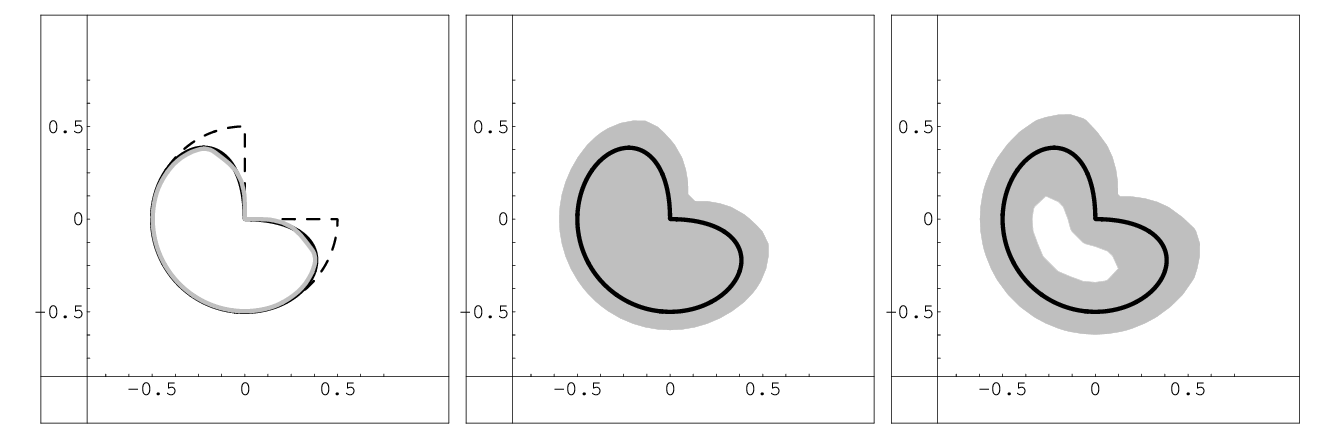}
\end{center}
\caption{
Left: the expected boundary $\Gamma[{\rcs A}]$ (black),  its estimate based on 25 samples (grey) and the boundary of a pacman with radius 0.5 (dashed); centre and right:  95\% bootstrapped confidence regions for $E[\rcs A]$ and $\Gamma[\rcs A]$, respectively.  The confidence sets are denoted by the shaded area, while the black line shows $\Gamma[\rcs A]$.  }
\label{fig:CIpacman}
\end{figure*}

\begin{example}[pacman in $\RR^2$]\label{ex:pacman}
Define the pacman with radius $r$, $A(r)$, to be a disc with radius $r$ centred at the origin with its upper left quadrant removed.  That is,
$$
A(r)=\{x:  |x|\leq r\} \cap \left\{\{x:  x_1 \leq 0 \} \cup \{x:  x_2 \leq 0 \}\right\}.
$$
Figure~\ref{fig:CIpacman} (right, dashed) shows the contour of $A(r)$ for $r=0.5$. 
Suppose that  $\rcs A=\rcs A(R),$ where $R$ is a uniform random variable on $[0,1].$ 

 The expected set $E[\rcs A]$ is a smoothed version of $A(0.5)$, as seen in
Figure \ref{fig:CIpacman} (left).   Recall that the smoothness of the boundary depends on the smoothness of $E[b_{\rcs A}(x)]$.  As the latter is an integral, which tends to have more smoothness than the original function, we expect that in general the expectation is as or more smooth than the original realizations of the boundary.  This is exactly what one sees in Figure \ref{fig:CIpacman}:  the mean boundary is smoothed out in the regions where there is movement; however, since the origin is a fixed point of the random boundary, no additional smoothness is introduced here by the average.

Figure \ref{fig:CIpacman} also shows bootstrapped $95\%$ confidence sets for both $E[{\rcs A}]$ and $\Gamma[{\rcs A}].$  
The accuracy of the estimate and the apparent centering of the confidence intervals around the mean set may be explained upon closer inspection of the sample.  The ODF of the pacman is similar to that of the circle;  indeed, they are identical in the lower left quadrant.  Therefore, the behaviour of the estimators and of the confidence regions is not unlike that of estimators and confidence intervals for the real-valued $E[R]=0.5$.  For our sample of $n=25$, we observed $\bar R_n=0.496,$ which explains the accuracy of the estimator $\overline{\rcs \Gamma}_n$.  On the other hand, the confidence region still shows the large variability of $\overline{\rcs \Gamma}_n$.\\
\end{example}

\noindent \textbf{RCS with simplified ODF structure.}

\smallskip

\noindent Consider an RCS $\rcs A$ and suppose that there exist functions $h_j$ $j=1, \ldots, k$ and random variables $\eta_j$ $j=1, \ldots, k$ such that 
\begin{eqnarray*}
&b_{\rcs A}(x) = \sum_{j=1}^k h_j(x) \, \eta_j&
\end{eqnarray*}
for all $x\in\mc D$ (see also Section 3 in \cite{stanberry:j:08}).  For example, the ball centered at $x_0$ with random radius $R$ takes this form.  Here,  $b_{\rcs{A}}(x)=|x-x_0|-R$, and hence $h_1(x)=|x-x_0|, h_2(x)=-1$ and $\eta_1=1, \eta_2=R.$

For such sets, both the empirical and ODF average have a similar simplified form.  That is,  $\bar b_n = \sum_{j=1}^k h_j(x)\, \bar \eta_{n,j}$ and $E[b_{\rcs A}(x)] = \sum_{j=1}^k h_j(x) \, E[\eta_j].$  Furthermore, the fluctuation field for these sets is particularly straightforward.  Suppose that we observe IID samples of the random vector $(\eta_1, \ldots, \eta_k)$, 
and assume that  $E[\eta_j^2]<\infty$ for all $j=1, \ldots, k$.  Then 
\begin{eqnarray*}
&\ZZ_n(x)
\Rightarrow \sum_{j=1}^k Z_j h_j(x),&
\end{eqnarray*}
where $Z=\{Z_1, \ldots, Z_k\}$ is a multivariate normal random variable with mean zero and variance matrix given by $\cov(Z_j, Z_m)=\cov(\eta_j, \eta_m).$

\begin{example}[Random half-plane]
Suppose that $\Theta\sim$Uniform$[a,b]$ and let $\rcs A = \{x: x_2\geq x_1 \tan\Theta\}.$  That is, $\rcs A$ is the plane above the line which goes through the origin and has angle $\Theta$ with the $x_1$-axis.  Here, $b_{\rcs A}(x)=x_1 \sin \Theta - x_2 \cos \Theta,$ and hence $h_1(x)=x_1, h_2(x)=x_2$ and $\eta_1=\sin\Theta, \eta_2=\cos\Theta.$  Some calculations also reveal that $E[\rcs A]=\{x: x_2\geq x_1 \tan((a+b)/2)\}$ and   $\Gamma[\rcs A]=\{x: x_2= x_1 \tan((a+b)/2)\}.$

The confidence region for $\Gamma[\rcs A]$ is a strip centred on the line $x_2=x_1 \bar \eta_1/\bar \eta_2$ of width $q_2/\sqrt{n}\, \bar \eta_2, $ where $q_2$ is the $1-\alpha$ quantile of $\sup_{x\in D} |Z_1 x_1+Z_2 x_2|.$
\end{example}

\subsection{Equivariance Properties}

The following proposition gives equivariance properties of the confidence regions under dilation and rigid motion.  The result corresponds to the classical scaling results for the mean and standard deviation of univariate data.

\begin{proposition}\label{prop:equi}
Consider a random closed set $\rcs A \subset \mc D.$  Let $\rcs C$ and $\rcs C_{\Gamma}$ denote the confidence regions for $E[\rcs A]\cap \mc W$ and $\Gamma[\rcs A]\cap \mc W,$ respectively.
\begin{enumerate}
\item Fix $\alpha>0$, and let $\rcs A_1 = \alpha \rcs A$ with $\mc W_1 = \alpha \mc W.$  Then the confidence regions for $E[\rcs A_1] \cap \mc W_1$ and $\Gamma[\rcs A_1] \cap \mc W_1$ are  $\alpha \rcs C$ and $\alpha \rcs C_{\Gamma},$ respectively.
\item Fix a rigid motion $g\in E(d)$, and let $\rcs A_2 = g(\rcs A)$ with $\mc W_2 = g(\mc W).$ Then the confidence regions for $E[\rcs A_2] \cap \mc W_2$ and $\Gamma[\rcs A_2] \cap \mc W_2$ are  $g(\rcs C)$ and $g(\rcs C_{\Gamma}),$ respectively.
\end{enumerate}
\end{proposition}

\subsection{A Modified Approach}\label{sec:mod}

It is not hard to see that the variability of $\bar {\rcs A}_n$ and  $\overline{\rcs\Gamma}_n$ depends on the  \emph{local} fluctuations of the field $\ZZ$ around $\Gamma[\rcs A]$ (see also \citet[Theorem 3.1]{molch:98}).  One natural way to incorporate this idea into calculation of the confidence region is described below.  
\begin{enumerate}
\item Calculate the $100(1-\alpha)\%$ confidence regions for $\Gamma[\rcs A]$ as described in the previous section.  Let $C$ denote this region.  
\item Find the $100(1-\alpha/2)\%$ upper quantile of $\sup_{x\in C} |\ZZ(x)|$, and call this~$\tilde q_2$.
\item The $100(1-\alpha)\%$ confidence region for $\Gamma[\rcs A]$  is then calculated as
$\left\{x: |\bar b_n(x)| \leq  \tilde q_2/\sqrt{n}\right\}.$
\end{enumerate}
The modification is designed to decrease the size of the quantile $q_1$ in \eqref{line:EACI}.  In the first step we reduce the size of the domain of the supremum to a set which is likely to contain the boundary.  Note also that the set $C$, and therefore also $\tilde q_2,$ depend on the sample size $n$.  Thus, the larger the sample size, the more effective the modification.   A similar approach yields a modified confidence region for $E[\rcs A].$    We find that this modification yields a slight improvement in the coverage probabilities for some settings, although visually the confidence regions remain quite similar.   Notably, iterating (i)--(iii) does not improve the size of the region or the coverage probabilities.

\section{Examples}\label{sec:examples}

\subsection{Implementation}

There exist several efficient algorithms to calculate the distance function of any set, which allows for easy implementation of our methods \citep{breu:95,freidman:77,rosenfeld:66}.  For many of the examples presented here, the oriented distance function may be calculated exactly.   When this was not possible, our calculations were implemented in MATLAB (MathWorks), where the \texttt{bwdist} command was used to compute the oriented distance function of a set.  Here, the examples need to be discretized to pixels (in $\RR^2$) or voxels (in $\RR^3$).    
This discretization introduces an additional source of error;  see \cite{serra:84} for a thorough treatment of the induced difficulties.  

To minimize the effect of discretization, in the simulations described below, we selected a fine gird, which was calibrated to give accurate results.  Suppose that $\mc D \subset \RR^2$, and that we observe $n$ discs centered at the origin with random radius $U\sim$Uniform[0,1].  Here, the confidence regions for the mean boundary are exact (modulo the sample size approximations).  Setting $n=1000$, we obtained empirical coverage probabilities of 95.10, 95.02, 95.30, 95.30, and 94.96 for the 95\% confidence regions, for grids with side length $m=200, 400, 600, 800, 1000$, respectively (the standard error due to bootstrap sampling was 0.0031).  Finally, we selected $m=400$ for our simulations.


\subsection{Simulation Study}\label{sec:sims}

\begin{table*}[htb!]
\caption{Empirical coverage probabilities for the expected set / expected boundary.}  
\medskip
\centering
\begin{tabular}{ccccc}
\toprule[1.5pt]
& \multicolumn{2}{c}{$n=25$}&\multicolumn{2}{c}{$n=100$}\\
\cmidrule(l){2-3}\cmidrule(l){4-5}\\
$100(1-\alpha)\%$ &$ 90\%$ & 95\% & 90\% & 95\%\\
\midrule
{(A)}    & 88.40/89.65     & 94.76/95.60   & 90.40/91.23      & 95.68/94.12   \\
{(B)}    & 90.24/89.85     & 94.63/95.05   &   90.16/90.35    & 95.14/95.04   \\ 
{(C1)}   & 90.07/91.15      & 94.36/95.10   & 91.64/91.05      & 95.28/95.29  \\
{(C2)}   & 91.39/93.49      & 96.45/97.37   & 92.14/93.31      & 96.85/97.13\\
{(D1)}  & 91.99/90.98      & 96.32/95.73    &  91.81/91.70& 96.20/95.74\\
{(D2)} &  90.67/88.56    &  94.48/94.66   &  90.95/88.89 &  94.86/94.97\\
\bottomrule[1.5pt]
\end{tabular}
\label{table}
\end{table*}

\begin{figure}[htb!]
\centerline{\fbox{\includegraphics [width=0.45\textwidth]{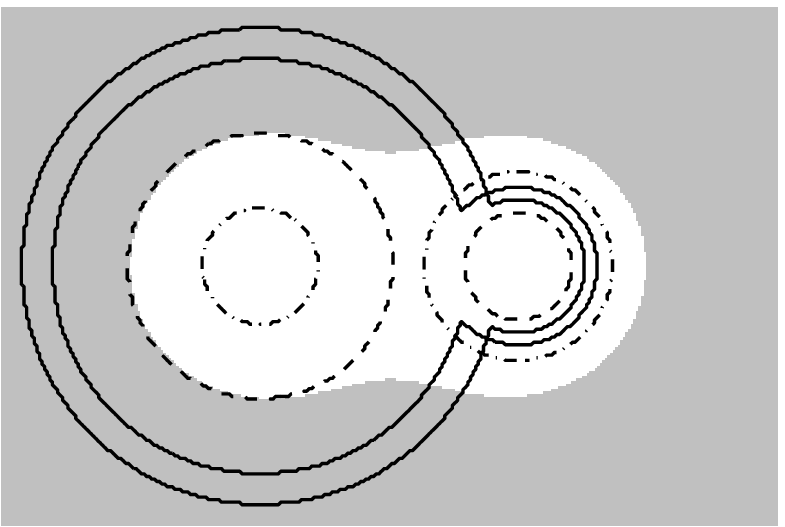}}}
\caption{The mean set in Example (C2) in white on a gray background.  The boundaries of several sample sets are also shown.}
 \label{fig:c2}
\end{figure}

We next simulate coverage probabilities for several examples of random closed set models.  The particular examples we consider are given below.
\begin{enumerate}
\item[(A)]  The set and its boundary when $p=0.5$, considered in Example \ref{ex:setboundary}.   Recall that in this example netiher the set estimator nor the boundary estimator is consistent.  Here $\mc W=[-1,2].$
\item[(B)]  The pacman with random radius $R\sim$Uniform[0,2] (see Example \ref{ex:pacman}).  Here $\mc W=[-2,2]^2.$  Although the pacman RCS is not decomposable, it still exhibits similar behaviour.
\item[(C)]  The random set $\rcs A= \rcs A_1 \cup \rcs A_2,$ where $\rcs A_1$ and $\rcs A_2$ are both random discs.
The two cases we consider are as follows.
\begin{enumerate}
 \item[(1)] $\rcs A_1$ is centered at the origin and has radius $R_1\sim$Uniform[0,2].    $\rcs A_2$ is centered at the point $(3,0)$ and has radius $R_2\sim$Uniform[0,1].  Here $\mc W=[-2,4]\times[-2,2].$
 \item[(2)] $\rcs A_1$ is centered at the origin and has radius  $R_1\sim$Uniform[0,1].    $\rcs A_2$ is centered at the point  $(1,0)$ and has radius $R_2\sim$Uniform[0,1].  Here  $\mc W=[-1,2]\times[-1,1].$  The mean set and some sample sets are shown in Figure \ref{fig:c2}.
 \end{enumerate}
\item[(D)]  A random ellipse with boundary parameterized as  $(x_1/R_1)^2+(x_2/R_2)^2=1$.  
Let  $U_1, U_2$ be two independent random variables with distribution  Uniform$[0,1].$  
The two cases we consider are as follows.  Throughout, we assume that $\mc W=[-1.5,1.5]\times[-1,1].$ 
\begin{enumerate}
 \item[(1)] $R_1 = 1, R_2=0.5+U_2/2$.
 \item[(2)] $R_1 = 1+U_1/5, R_2=0.5+U_2/2$.
\end{enumerate}
\end{enumerate}

In the above examples, the image was discretized and the quantiles were estimated using Monte Carlo methods (with $B=2000$ repetitions and $n=100$), except for example (A), where the quantile is known exactly.  The empirical coverage was estimated using $10,000$  Monte Carlo simulations in each case.  It is important to note that there are four sources of error: sample size, discretization, the bootstrap estimation of the quantile, and the \linebreak Monte Carlo simulations themselves.  The maximal standard error due to Monte Carlo of the empirical coverage probabilities is $0.30\%$, but it is not easy to quantify the standard error due to the other three sources.  In particular, we note that the quantile was estimated once per example, and the same quantile was used in each Monte Carlo simulation (otherwise, the simulations would have been prohibitive), which increases the bias of our results.

The results of the simulation study show that our methods, though conservative, achieve good coverage probabilities.  The greatest over-coverage is seen in Example (C2), which is the most difficult of the examples because the observed sets have not been co-registered;  see Figure \ref{fig:c2} for some sample sets.    In this example, it is possible that a slight improvement of the coverage probability could be seen by a modification to the confidence region discussed in Section \ref{sec:mod}.  In general, to minimise over-coverage, we recommend choosing $\mc W$ as small as possible in practice.

\subsection{A Toy Image Reconstruction Example}

\begin{figure}[t!]
\centerline{\fbox{\includegraphics [width=0.45\textwidth]{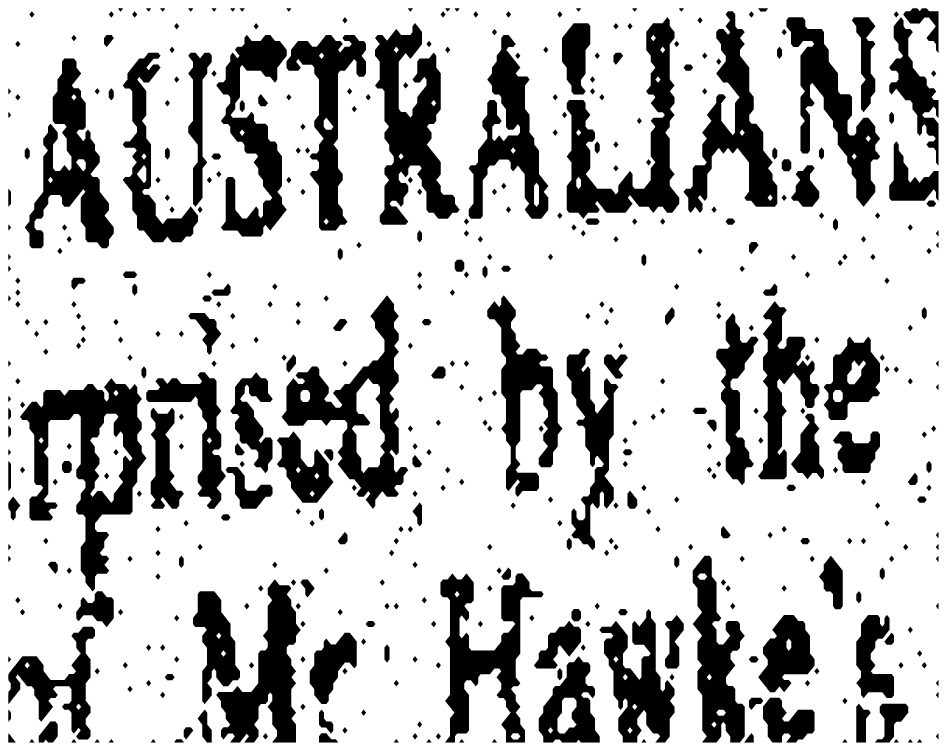}}
\fbox{\includegraphics [width=0.45\textwidth]{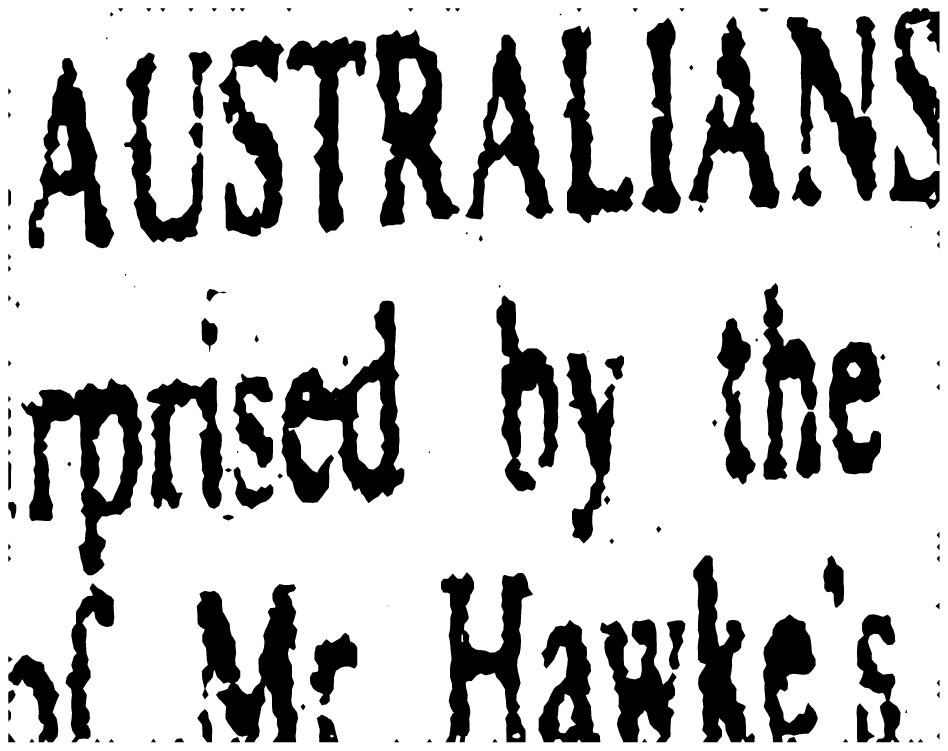}}}
\caption{A sample reconstructed image (left) and ODF average based on 15 IID samples from the posterior (right).}
 \label{fig:newsprint1}
\end{figure}

Image averaging arises in various situations, for example, when multiple images of the same scene are observed or when the acquired images represent objects of the same class and the goal is to determine the average object (shape) that can be described as typical.  Here we consider an example of image averaging studied in \cite{baddeley:molchanov:98}.  The data set consists of 15 independent samples of a reconstructed newspaper image (Figure \ref{fig:newsprint1},  left), and is available from \texttt{http://school.maths.uwa.edu.au/homepages/adrian}.   For details on how the data was generated we refer to \cite{baddeley:molchanov:98}.  

\begin{figure}[b!]
\centerline{\fbox{\includegraphics [width=0.45\textwidth]{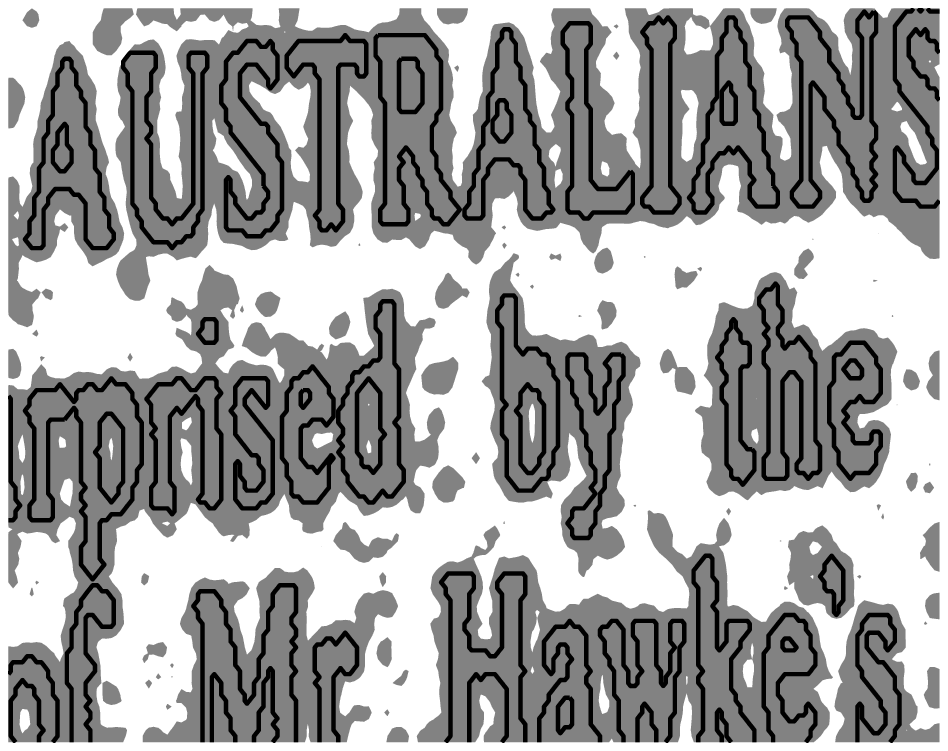}}
\fbox{\includegraphics [width=0.45\textwidth]{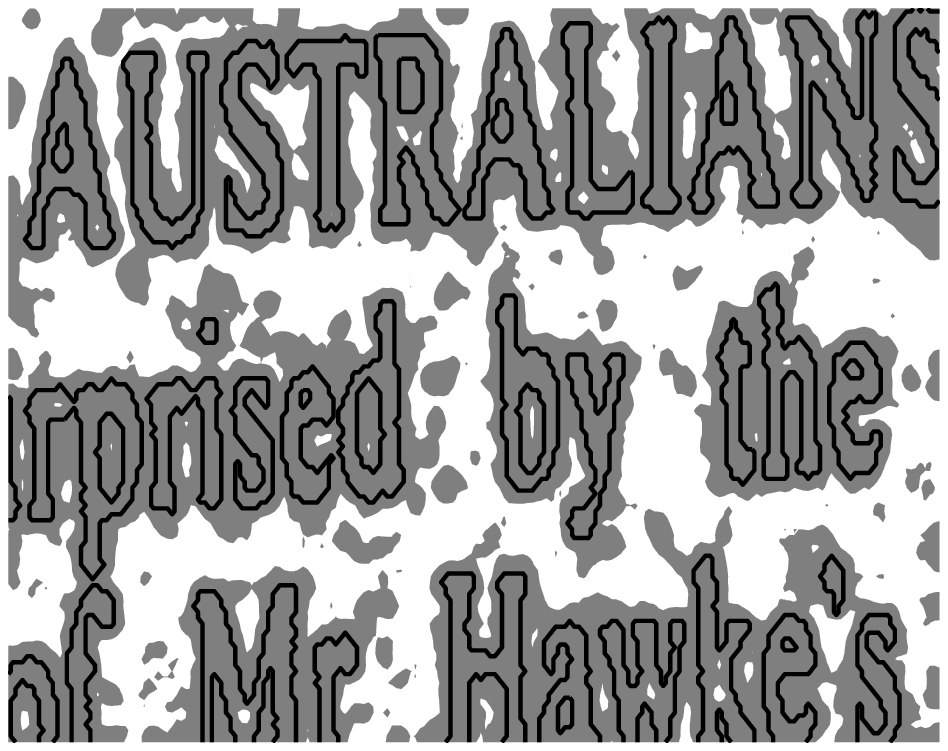}}}
\caption{95\% confidence regions for $E[\rcs A]$ (left) and $\Gamma[\rcs A]$ (right).  The boundary of the original newspaper image is shown in black.}  \label{fig:newsprint2}
\end{figure}

\begin{figure*}[t!]
\centerline{
    \fbox{\includegraphics [width=0.31\textwidth]{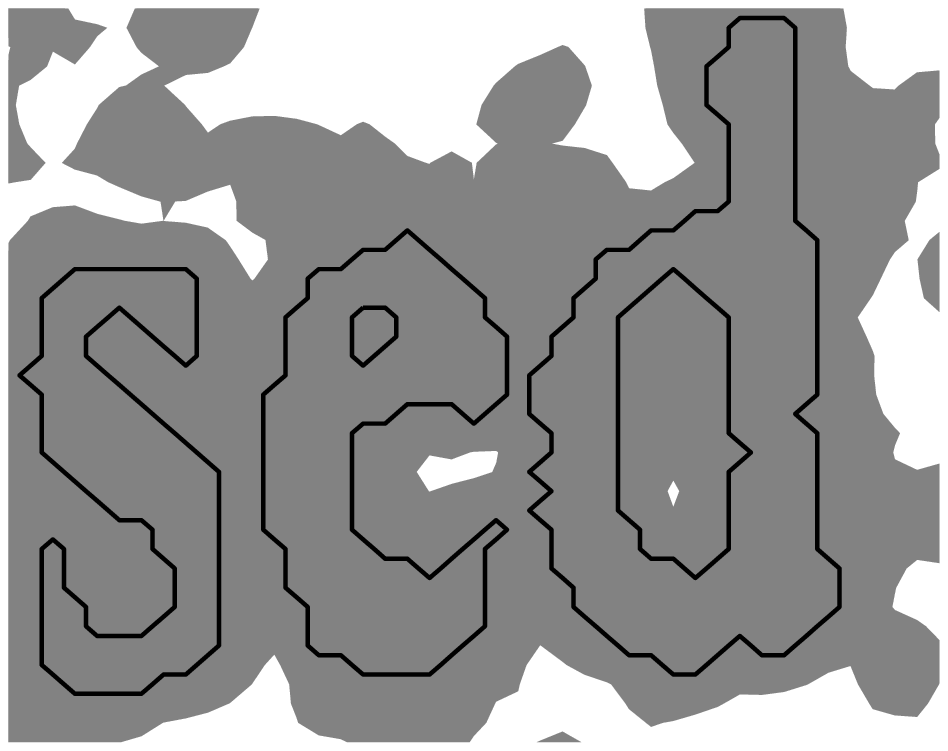}}
    \fbox{\includegraphics [width=0.31\textwidth]{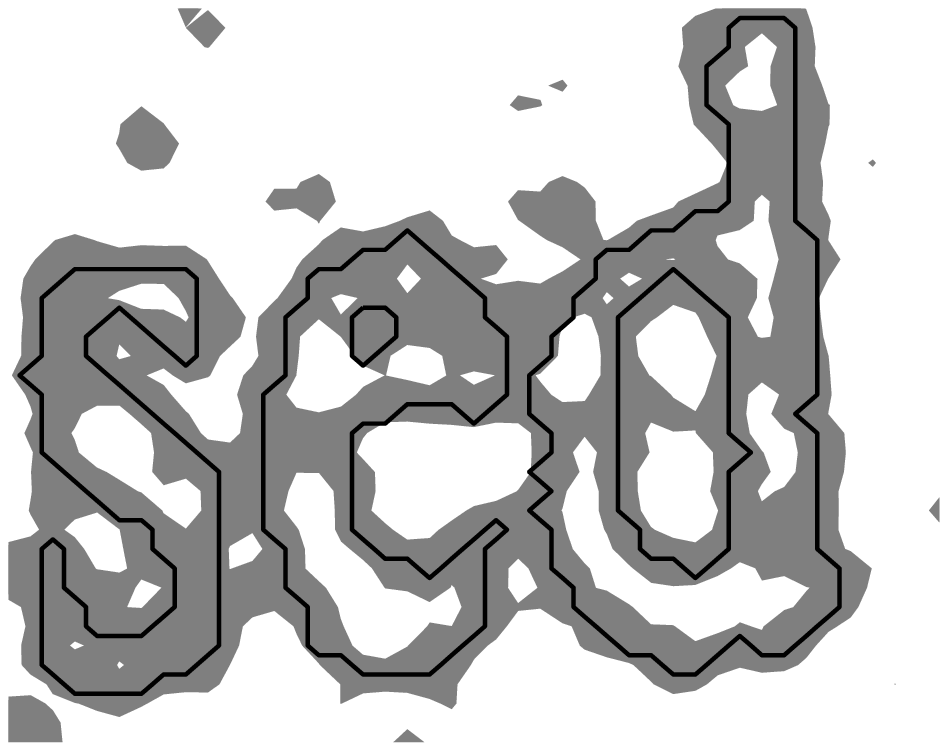}}
    \fbox{\includegraphics [width=0.31\textwidth]{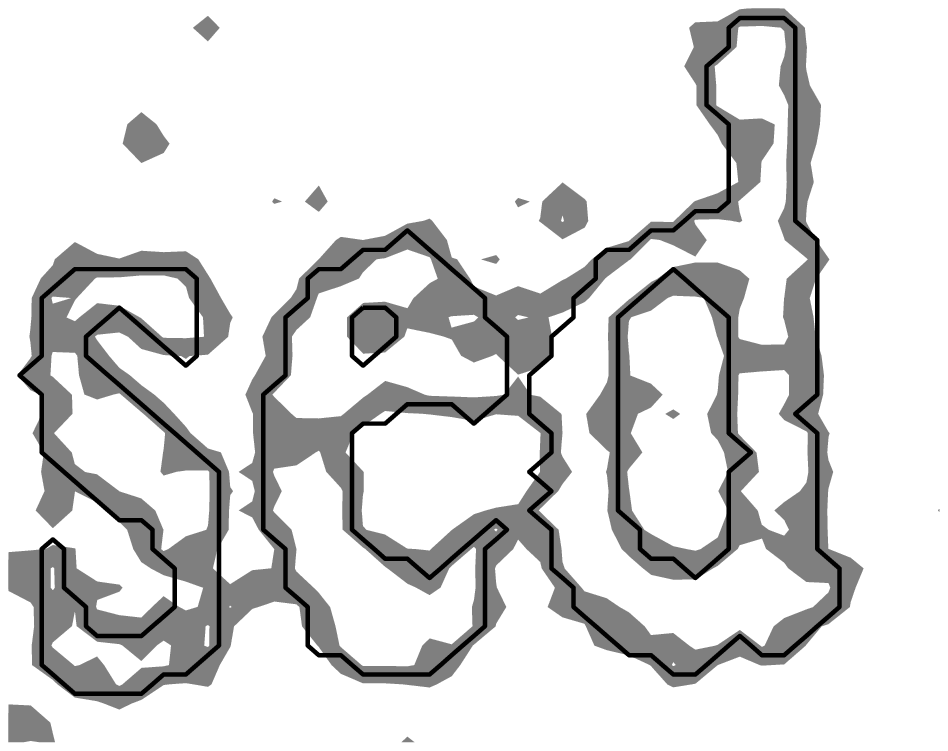}}
    }
\caption{Confidence regions for the expected boundary with the boundary of the true image (black) using the true sample size (left) and hypothetical sample sizes of $n=50, 100$ (middle and right, respectively). The pictures shown are insets of the full image.}\label{fig:odfbEAlargen}
\end{figure*}

The empirical average of the 15 observed images, $\bar{\rcs A}_n$ is shown in Figure \ref{fig:newsprint1} (right).  The average $\bar{\rcs A}_n$ describes the ``typical" image reconstruction, and as such, may be thought of as an estimator of the true text image. 
Next, we compute 95\% confidence regions for the ODF-average reconstruction based on 5K bootstrap samples.  We may think of these regions as a measure of the variability of $\bar{\rcs A}_n$, and $\overline{\rcs \Gamma}_n$.  
Figure \ref{fig:newsprint2} (top) shows the confidence region for $E[\rcs A]$ with the boundary of the true image overlayed in black. The confidence set contains all of the true text image and appears tight, although there are a number of spurious bounds induced by noise.  The confidence set for $\Gamma[\rcs A]$ is also shown in Figure \ref{fig:newsprint2} (right).  The boundary of the lower confidence set consists of only a few closed contours scattered throughout the text. The lack of tightness in the confidence set is explained by the small sample size and a relatively thin font width.  Hypothetically increasing the sample size would produce tighter confidence sets for both the expected set and its boundary. For example, the bootstrap confidence set for the boundary based on 50 and 100 samples is tighter as compared to the one based on 15 samples (see Figure \ref{fig:odfbEAlargen}).  We expect that actual confidence regions for $n=50$ and $n=100$ would exhibit even less noise than the hypothetical regions shown in Figure \ref{fig:odfbEAlargen}.  It should be noted that the confidence regions in Figure \ref{fig:odfbEAlargen} were created using the original window $\mc D$, and the picture is a close-up of the~result.

Note that these images were generated under a Bayesian framework, where the goal is to reconstruct the original image by selecting an appropriate parameter from the posterior distribution.  Here, this parameter is not computable directly, and is instead estimated by $\bar{\rcs A}_n$ and $\overline{\rcs \Gamma}_n$.  Although a frequentist concept, the confidence regions allow us to determine the variability of these estimators.  As Figure \ref{fig:odfbEAlargen} shows, this variability depends on the sample size.  

\subsection{Analysis of Sand Grains}

\begin{figure}[b!]
\centerline{
\includegraphics[scale=0.4]{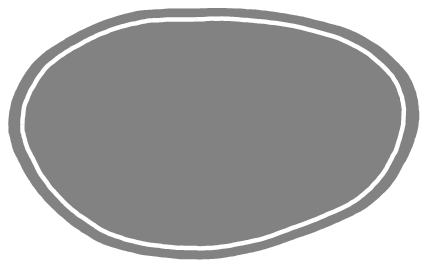} \hspace{10mm}
\includegraphics[scale=0.4]{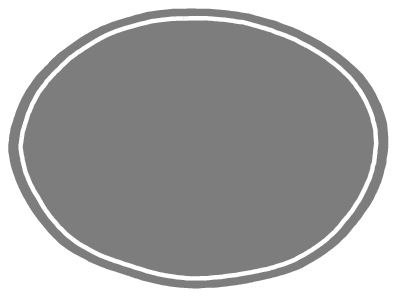}
}
\vspace{5mm}
\centerline{
\includegraphics[scale=0.4]{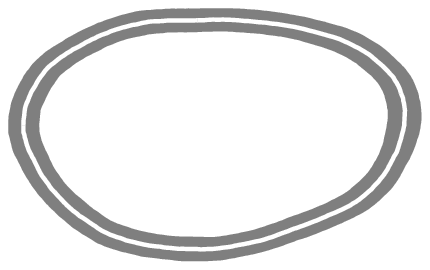}\hspace{10mm}
\includegraphics[scale=0.4]{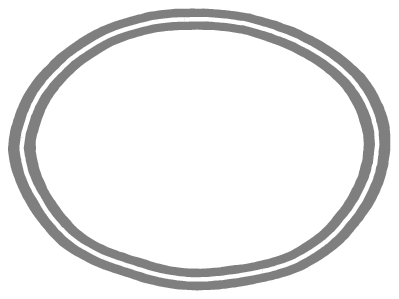}
}
\caption{Confidence regions (grey) for the average grain (top row) and the average boundary (bottom row) from the Zelenchuk river (left column) and the Baltic Sea (right column), based on realignment with scaling.}\label{fig:grainsCI}
\end{figure}

Next, we apply the proposed methods to the sand grains data previously described in \cite{stoyan:97} and \cite{kent:00}. The sand particles were collected from the shores of the Baltic Sea and banks of the Zelenchuk River in Ossetia. The grains were photographed on the same scale and the data resembles two dimensional projections represented by binary images.  Both images may be found online at \texttt{www.math.yorku.ca\\/$\sim$hkj/Research/SandGrains} (please note that the online images are not to scale).   The observed sand grains generally have a smooth rounded shape, but are not necessarily covex.  

The grains from the two regions appear to differ both in shape and size with sea grains being more spherical and larger as compared to river grains that are more oblong and smaller in diameter.   To summarize the data, we find the average grains and their average boundaries for each group.  We also construct confidence regions to describe the variability of the grain shapes.

\begin{figure}[b!]
\centerline{
\includegraphics[scale=0.4]{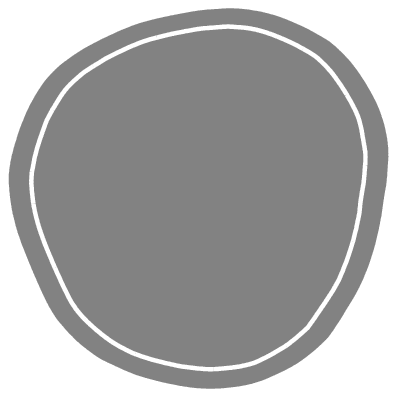} \hspace{10mm}
\includegraphics[scale=0.4]{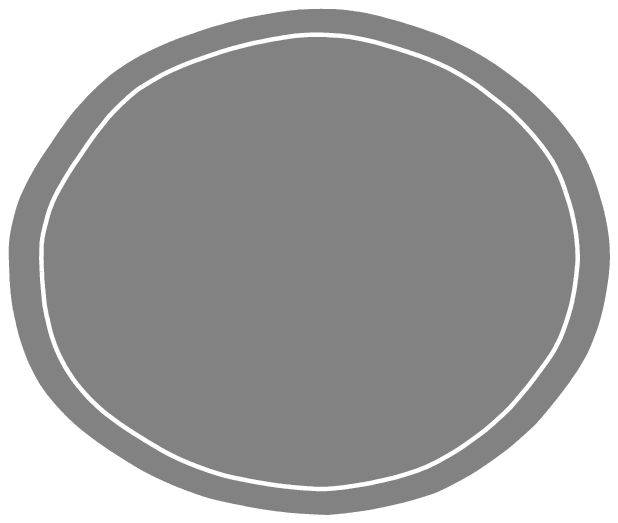}
}
\vspace{5mm}
\centerline{
\includegraphics[scale=0.4]{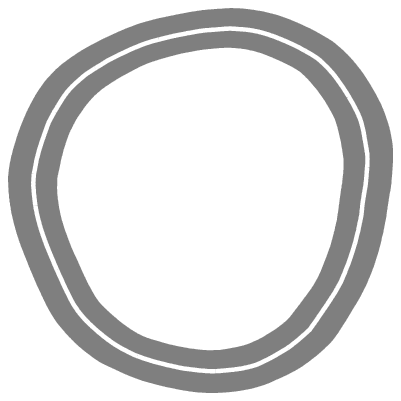}\hspace{10mm}
\includegraphics[scale=0.4]{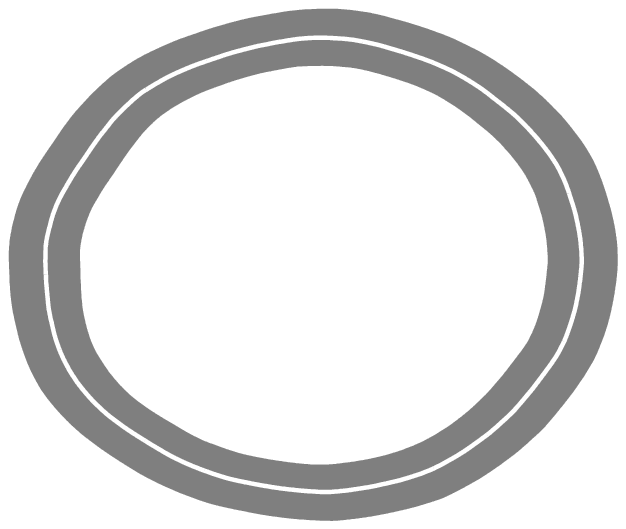}
}
\caption{Confidence regions (grey) for the average grain (top row) and the average boundary (bottom row) from the Zelenchuk river (left column) and the Baltic Sea (right column), based on realignment without scaling.}\label{fig:grainsCInoscale}
\end{figure}

To begin with, the particles were realigned using the generalised Procrustes analysis as implemented in the {\tt shapes} package in R \citep{R}.   To apply the Procrustes analysis, we use the digitized data as described by \cite{kent:00}.
The data was digitized so that each sand grain was represented by 50 vertices approximately equally spaced on the boundary.  After digitization, the arc-length between the vertices is around 10-20 pixels with grain particles represented by high-resolution images of size $500 \times 350$.   The realignment was done with and without scaling. Using the scaling, we essentially remove the size effect and can examine differences in average shapes. Alternatively, average particles based on Procrustes analysis without scaling reflect differences both in size and shapes of the particles. The median (IQR) centroid sizes are 1481 (1396, 1665) and 2076 (1867, 2376) for the river and sea grains, respectively, indicating the sea particles to be bigger as compared to the river ones.

Figure \ref{fig:grainsCI} shows the confidence regions for the average particle (top row) and the expected boundary (bottom row) for the river (left column) and the sea (right column) sands using scaled realignment. White contours show the empirical mean boundary. The confidence regions are based on 5K bootstrap samples. The average river grain is more oblong as compared to the sea grain. The variability within the two groups appears to be rather similar.   Figure \ref{fig:grainsCInoscale} shows the confidence regions for the average particle (top row) and the expected boundary (bottom row) for the river (left column) and the sea (right column) sands, based on realignment without scaling. White contours show the empirical mean boundary. The confidence regions are based on 5K bootstrap samples. The average sea grain is more spherical in shape as compared to the river average. It is also considerably larger in size. The variability within the two groups again appears to be rather similar, however, the unscaled images appear more variable than the scaled  ones.   Overall, the discrepancies in shape and size between the two averages reflect the differences between the raw data sets.   Note that the boundaries of the average sets in both figures are rather smooth.

There is also a marked difference between the scaled and unscaled river sand grain averages.  Image results from the Procrustes analysis re--alignment with and without re--scaling are quite different. 
This is to be expected, as the scaling, location, and centering re--alignments in Procrustes analysis are highly interdependent.   It would be of interest to compare other re--alignment methods, such as those proposed by \cite{stoyan:97}, but this is beyond the scope of this work.

Whether or not scaled Procrustes realignment is applied, the averages of the particles show a clear difference between the two groups.     However, this difference is at this time only visual.  An interesting and important problem is to develop quantifiable methodology to test for presence and locations of differences between the mean shapes.

\subsection{Application to Medical Imaging}

\begin{figure*}[htb!]
\centerline{
\begin{tabular}{ccc}
\multirow{3}{*}{\rotatebox{-90}{{\includegraphics [width=0.45\textheight, height=0.35\textwidth]{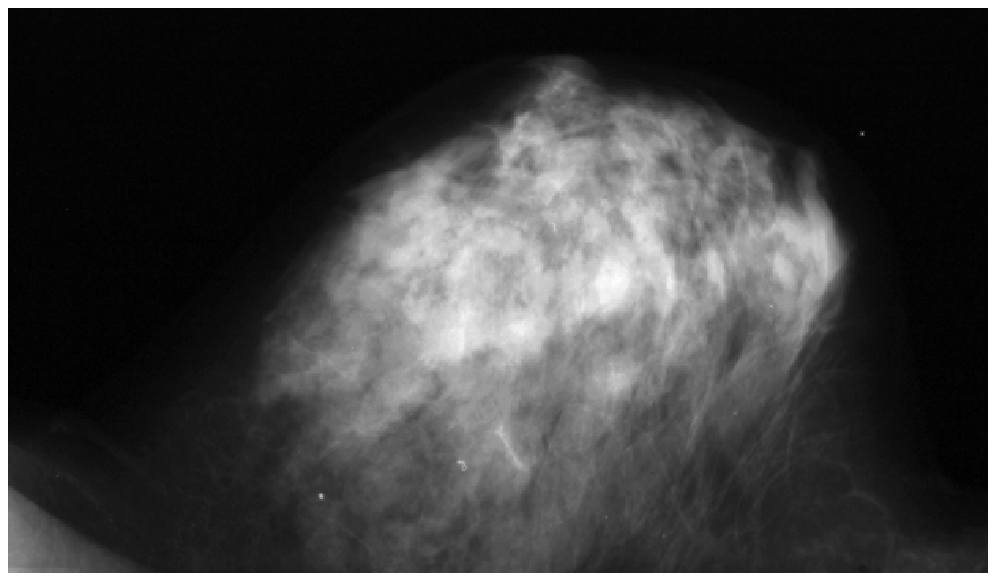}}}} &
\multirow{3}{*}{\rotatebox{-90}{{\includegraphics [width=0.45\textheight, height=0.33\textwidth]{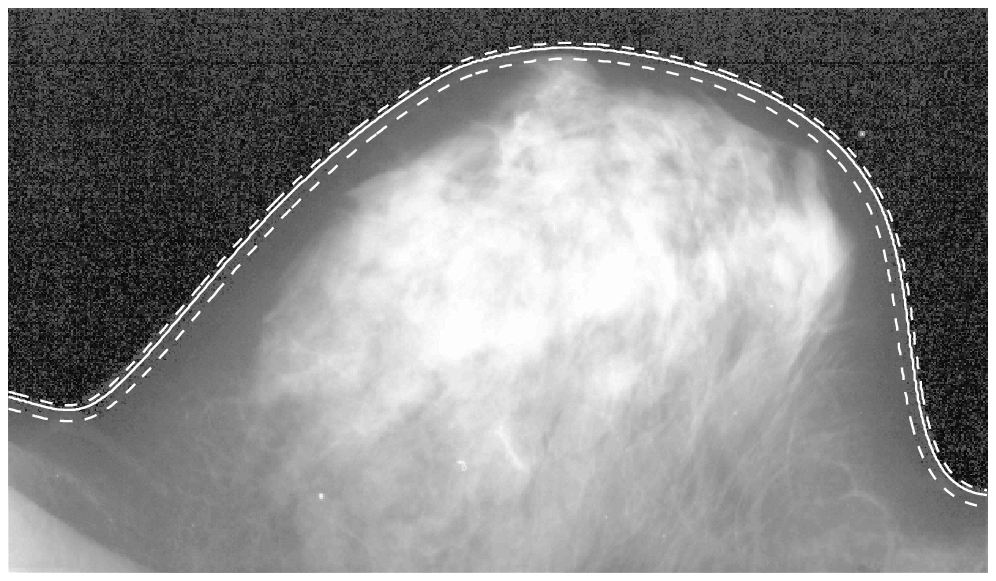}}}} &
\rotatebox{-90}{{\includegraphics [width=0.13\textheight, height=0.22\textwidth]{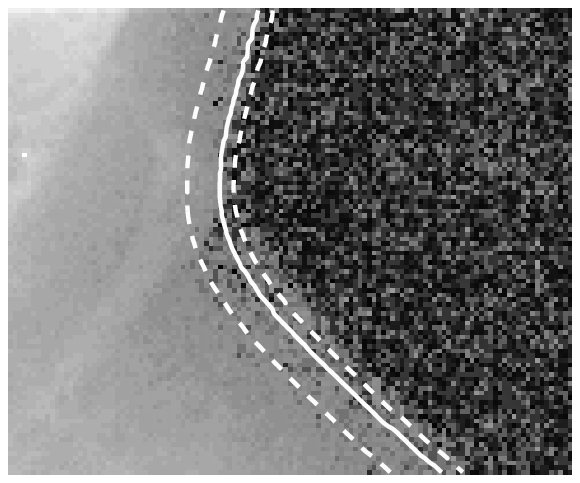}}}\\
 & &  \rotatebox{-90}{{\includegraphics [width=0.164\textheight, height=0.22\textwidth]{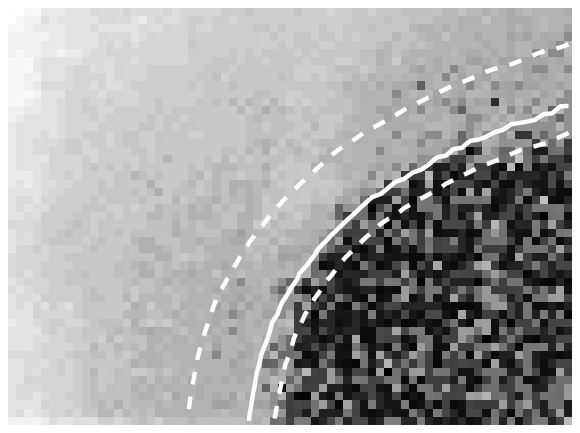}}}\\
 & & \rotatebox{-90}{{\includegraphics [width=0.13\textheight, height=0.22\textwidth]{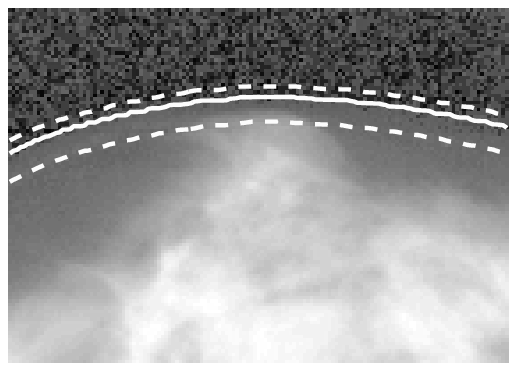}}}\\\\
\end{tabular}
}
\caption{Confidence sets for the reconstructed skin-air boundary in a mammogram:  the original image (left), and the digitally enhanced image (centre) with the reconstructed boundary (solid line) and confidence region (dashed line).  Three insets are also shown (right).}
\label{fig:mgram}
\end{figure*}

We next consider an example of boundary reconstruction in mammography, where the skin-air contour is used to determine the radiographic density of the tissue and to estimate breast asymmetry.  Both measures are known to be associated with the risk of developing breast cancer \citep{scutt:97, ding:2008}.  In \cite{stanberry:besag:08}, B-spline curves were used to reconstruct a smooth
connected boundary of an object in a noisy image, and a Bayesian approach was applied
to estimate the tissue boundary in mammograms.

The boundary reconstruction was performed on a binary image which was obtained  after filtering and thresholding the original greyscale mammogram.   Let $\mc D$ be the compact domain of the mammogram image, and let $\rcs T$ denote the random set describing the breast tissue, or foreground, of the image under the prior belief distribution.    Next, let $M$ denote the noisy binary mammogram image observed.   The skin-air boundary estimate is reconstructed as $\Gamma[\rcs T|M]$, the expected boundary of $\rcs T$ from the posterior distribution given the observed data $M$.     The posterior distribution of the random set $\rcs T|M$ is too difficult to compute, and was approximated via Markov chain Monte Carlo.  Hence, the skin-air boundary estimate $\Gamma[\rcs T|M]$ was also approximated as $\overline{\Gamma}_n$
from a sub-sample of observed random sets $T_i, T_{i+1}, \ldots$ generated from the MCMC simulation.  Further details can be found in \cite{stanberry:besag:08}.  
Here, we apply the proposed method to construct a confidence region for the posterior mean boundary.  
We emphasize that the confidence region is not a credible set, but rather it describes the variability of $\overline{\Gamma}_n$ as an estimator of $\Gamma[\rcs T|M].$

Figure \ref{fig:mgram} (left) shows a typical digitized mammogram image, characterised by a low contrast-to-noise ratio. A probability integral transform improves the contrast by increasing the dynamic range of image intensities (Figure \ref{fig:mgram}, centre).   The solid white line in Figure 10 (centre and insets on right) shows the reconstructed boundary $\overline{\Gamma}_n.$ Note that what appears to be a nipple is, in fact, a duct system leading to the nipple, so that the estimators correctly follow the skin line.

The 95\% confidence set (dashed) for the true boundary in Figure \ref{fig:mgram}
is obtained using a bootstrap resampling of size 1000. The
confidence set is tight and fits the image well. It also shows that the reconstructed boundary is more variable toward the inside of the breast tissue.  The
background of the black and white mammogram image has considerably more
noise than the foreground. Consequently, the posterior boundary samples show
more variability toward the inside of the tissue.   More details can be seen
in insets in Figure \ref{fig:mgram} (right).

Note also that to apply our methods, we have assumed that the observed ODFs $b_{T_i}(x)$ are independent and identically distributed, whilst the boundary
reconstruction is based on Markov chain Monte Carlo sampling from the posterior. To
ensure the independence of the curve samples, we construct the
confidence set for the boundary using 100 samples from the posterior,
which were acquired every 250th sweep after a burn-in period of 1000
sweeps. 

\section{Discussion}

In this paper, we studied consistency of set and boundary averages under random sampling.  We also presented a method for the construction of confidence regions for the mean set and mean boundary.  The confidence regions, though conservative, have appealing equivariance properties and are straightforward to implement.   Simulations indicate that they achieve good coverage probabilities.  Unlike previous developments, our methods are applicable to both convex and non-convex sets and allow for differences in local variability.  

As there exists no notion of the standard deviation of a random set, one can also use the confidence regions as an informal assessment of the variability of the mean set estimator.   This relationship is strengthened by the aforementioned equivariance properties, which mimic the scaling properties of the standard deviation and confidence region in the univariate setting.  

In Section 5 we considered several empirical examples.  The observed sets in these cases are non-convex, and therefore methods based on the Aumann expectation would not work well, although they may yield reasonable approximations for the sand grains example.    In Section 5.5 we applied the proposed methods to a boundary reconstruction problem in a mammogram image.    There, the confidence region technique was used to assess the variability of the MCMC estimation of the posterior mean.  The proposed method was able to detect an increase in the variability of the sampling toward the inside of the breast tissue.  A dilation method, such as one based on the results of \cite{molch:98}, would not be able to detect this difference.

\section*{Acknowledgements}
The first author would like to thank Tom Salisbury from York University for several useful discussions.  Both authors thank Ian Dryden from the University of South Carolina and Dietrich Stoyan from the Freiberg University of Mining and Technology for the sand grains data and for advice on the Procrusters transformations.  

\def\cprime{$'$}

\newpage


\renewcommand{\theequation}{A-\arabic{equation}}
\setcounter{equation}{0}  
\setcounter{section}{6}  

\section*{Appendix}

Recall that the boundary of a set $A\subset \mathbb R^d$ is $C_k$ in a neighbourhood $N(x_0)$ if there exists a bijective map $m:N(x_0) \mapsto B_1(0)$, which is $C^k$ and whose inverse is also $C^k$, which maps the boundary into the set $\{x\in B_1(0): x_d=0\}.$  That is, the boundary is $C^k$ at $x_0$ if locally it is a $C^k$ manifold.   

\begin{proof}[Proof of Proposition \ref{prop:smoothboundary}]

We first note that the condition $|\nabla E[b_{\rcs A}(x_0)]|\neq 0$ is necessary.  For example, if $E[b_{\rcs A}(x)]$ has a local minimum at $x_0$ then the boundary of $E[\rcs A]$ contains the isolated point $\{x_0\}$, and no smoothness properties may be carried from the expected oriented distance function to the expected boundary.


Write $x=(x_1, \ldots, x_d)\in \RR^d$ with $x_0=(x_{0,1}, \ldots, x_{0,d}).$   Since $|\nabla E[b_{\rcs A}(x_0)]|\neq 0$, there exists a $j$ such that $\partial_j E[b_{\rcs A}(x_0)]\neq 0.$   Let  $x_{(j)}=(x_1, \ldots, x_{j-1}, x_{j+1}, \ldots, x_d)\in \mathbb R^{d-1},$ and define $H:\mathbb R^{d-1}\times \bb R \mapsto \bb R$ by $H(x_{(j)}, x_j)=E[b_{\rcs A}(x)].$   The boundary is now described via the set $H(x_{(j)}, x_j)=0$ and we may apply the implicit function theorem (Theorem 1-12, page 41 of \cite{spivak} and Theorem 31 p. 299 of \cite{schwartz}).  The differentiability condition on the Jacobian in the implicit function theorem holds since
$$
\partial_{x_j} H (x_{(j)}, x_j) = \partial_j E[b_{\rcs A}(x_0)]\neq 0.
$$
It follows that there exists a function $g(x_{(j)})$, a neighbourhood of  \linebreak$(x_{1,0}, \ldots, x_{j-1,0}, x_{j+1,0}, \ldots, x_d)$, $N_1\subset \bb R^{d-1}$, and a neighbourhood of $x_{j,0}$, $N_2\subset\bb R$  such that $g:N_1\mapsto N_2$ is $C^k$ and describes the boundary, $\Gamma[\rcs A]$, near $x_0$.  
\end{proof}

\begin{proof}[Proof of Theorem \ref{thm:LLN}]
Pointwise convergence follows immediately by the law of large numbers. Since both $\bar b_n(x)$ and $E[b_{\rcs A}(x)]$ are Lipschitz functions \citep{stanberry:j:08}, we also obtain uniform convergence over compact sets.
\end{proof}

The next result may be found, for example, in \citet[Theorem 1.4.7]{kunita:90}.  We repeat it here for convenience.

\begin{theorem}[Kolmogorov's tightness criterion.]\label{thm:tight} For a compact set $\mc D\subset \RR^d$,
let $\{Y_n(x): x\in \mc D\}$  be a sequence of continuous random fields with values in $\RR$.  Assume that there exist positive constants $\gamma, C$ and $\alpha_1, \ldots, \alpha_d$ with $\sum_{i=1}^d \alpha_i^{-1}<1$ such that
\begin{eqnarray*}
E[|Y_n(x)-Y_n(y)|^\gamma] &\leq& C\left(\sum_{i=1}^d|x_i-y_i|^{\alpha_i}\right) \ \ \mbox{ for all }x,y\in \mc D,\\
E[|Y_n(x)|^\gamma]&\leq& C, \ \ \ \mbox{ for all }x\in \mc D,
\end{eqnarray*}
holds for any $n$.  Then $\{Y_n\}$ is tight in $C(\mc D)$.
\end{theorem}

\begin{lemma}\label{lem:clt} There exists a constant $C(d)$, depending only on $d$, such that
\begin{eqnarray*}
E\left[|\ZZ_n(x)-\ZZ_n(y)|^{2d}\right]&\leq& C(d)|x-y|^{2d},
\end{eqnarray*}
for any $n$ and $x,y \in \mc D.$
\end{lemma}

\begin{proof}
The case $d=1$ is immediate.  Next, consider $d=2$,
\begin{eqnarray*}
E\left[|\ZZ_n(x)-\ZZ_n(y)|^{4}\right]&=&n^{-2}\sum_{i,j,k,l=1}^n E\left[ b^*_ib^*_jb^*_kb^*_l\right],
\end{eqnarray*}
where $b^*_i = b_i(x)-b_i(y)-E[b_{\rcs A}(x)]+E[b_{\rcs A}(y)],$ and $|b^*_i|\leq 2 |x-y|$ almost surely, since both $b_i$ and $E[b_{\rcs A}]$ are Lipschitz (cf. \cite{stanberry:j:08}). Since the sampling is IID, and the $b^*_i$ are centred, it follows that the right-hand side of the above display is equal to
\begin{eqnarray*}
n^{-2} \left\{n E[(b_1^*)^4] + 3n(n-1) E[(b_1^*)^2]^2\right\} \leq 64 |x-y|^4.
\end{eqnarray*}
Similarly, for $d=3,$
\begin{eqnarray*}
&&E\left[|\ZZ_n(x)-\ZZ_n(y)|^6\right]\\
&=&n^{-3}\sum_{i,j,k,l,p,t=1}^n E\left[ b^*_ib^*_jb^*_kb^*_lb^*_pb^*_t\right]\\
&=& n^{-3} \left\{n E[(b_1^*)^6] \right.\\
&&\hspace{1cm}+3n(n-1)\left(E[(b_1^*)^3]^2+E[(b_1^*)^2]E[(b_1^*)^4]\right)\\
&&\left.\ \  \ \ \ \hspace{1cm} + \  90 n(n-1)(n-2) E[(b_1^*)^2]^3\right\}\\
&\leq& 97\cdot 2^6 \cdot |x-y|^6.
\end{eqnarray*}

In general, the expansion becomes
\begin{eqnarray*}
&&n^{-d}\left\{n E[(b_1^*)^{2d}]+ \ldots \right.\\
&&\left.\hspace{0.5cm}+ {2d \choose 2\  2 \ \ldots \ 2 } n(n-1)\ldots (n-d+1) E[(b_1^*)^2]^d\right\},
\end{eqnarray*}
which is bounded above by $C(d)|x-y|^{2d}$, for some constant $C(d).$
\end{proof}

\begin{proof}[Proof of Theorem \ref{thm:randomsetCLT}]
Recall that  $b_{\rcs A}(x)$ is almost surely Lipschitz.   Then $E[b_{\rcs A}(x_0)^2]<\infty$ for some $x_0\in \mc D,$ implies that $E[b_{\rcs A}(x)^2]<\infty$ for all $x\in \mc D$.  Therefore, convergence in finite dimensional distributions is immediate by the multidimensional central limit theorem, and it remains to prove that the process $\ZZ_n$ is tight in the space of continuous functions on $\mc D$. However, this is straightforward if we use Theorem \ref{thm:tight}.

The first condition with $\gamma=2d$ and $\alpha_i=2d$ for all $i$, follows immediately from Lemma \ref{lem:clt} by Jensen's inequality.  For the second condition we need to bound $E[\ZZ_n(x)^{2d}]$ uniformly.   This follows easily since, for some fixed $x_0\in \mc D$,
\begin{eqnarray*}
  E\left[\ZZ_n(x)^{2d}\right] \leq C' \left(E\left[\ZZ_n(x_0)^{2d}\right]+E\left[|\ZZ_n(x)-\ZZ_n(y)|^{2d}\right]\right)
\end{eqnarray*}
for come constant $C'$ (depending on $d$), again applying Jensen's inequality.  We have already placed a bound on the second term of the right-hand side of the above equation, and a bound on the first term follows from the central limit theorem.
\end{proof}

Let $\mc D$ be a compact subset of $\mathbb R^d.$  We recall a theorem of \cite{winkler:64}.  Proposition \ref{prop:lip} follows immediately.

\begin{theorem}[SATZ 6 on page 837 of \cite{winkler:64}]\label{thm:lipcond}
Let $\{Y(x), x\in  \mc D \subset \RR^d\}$ be a Gaussian random field such that for $\tau \rightarrow 0$ the inequality
$$
E\left[\left|Y(x+\tau)-Y(x)\right|^2\right] \leq C|\tau|^\varepsilon
$$
holds for some $\varepsilon >0$ and $0<C<\infty$.  Then for almost all realizations there exists a random number $\delta(\omega)$ so that for any  $x_1, x_2 \in  \mc D$ with $|x_1-x_2|<\delta(\omega)$ and $0<\eta<\varepsilon/2$ the inequality
$$
|Y(x_1)-Y(x_2)| \leq C_0 |x_1-x_2|^\eta
$$
holds.   In particular, it follows that $\{Y(x), x\in \mc D\}$  is continuous with probability one.
\end{theorem}

\begin{proof}[Proof of Proposition \ref{prop:lip}]
To prove this result we again recall that both $|b_{\rcs A}(x)-b_{\rcs A}(y)|\leq |x-y|$ almost surely.   Therefore,
\begin{eqnarray*}
  \var(\ZZ(x)-\ZZ(y))&=& \var(b_{\rcs A}(x)-b_{\rcs A}(y))\\
                     &\leq & E[(b_{\rcs A}(x)-b_{\rcs A}(y))^2] \leq |x-y|^2.
\end{eqnarray*}
A similar approach shows the bound for the covariance.  We may now use this result, along with Theorem \ref{thm:lipcond} to prove that the sample paths of $\ZZ$ are continuous almost surely.
\end{proof}

\begin{proof}[Proof of Theorem \ref{thm:consistent}]
The first part of the theorem follows directly from \citet[Theorem 2.1]{molch:98}.  The second part follows from \citet[Theorem 1]{cuevas:g:w:r:06}, but some further explanations are necessary.  Without loss of generality, we may assume that $\mc D$ is compact.  Therefore, note that (M1) and (f2)  of \citet[Theorem 1, page 9]{cuevas:g:w:r:06} are satisfied, and that the remaining condition (f1) holds under \eqref{cond:consistent_mean_A} and \eqref{cond:consistent_mean_B}.   Fix $\eps>0$.  To prove their result, they show that there exists an $n_0$ such that for all $n\geq n_0$
\begin{eqnarray*}
\partial\{x: E[b_{\rcs A}(x)]\geq 0\} &=& \{x: E[b_{\rcs A}(x)]= 0\} \\
&\subset & \left(\partial\{x:\bar b_n \geq 0\}\right)^\eps\\
&\subset & \{x:\bar b_n = 0\}^\eps,
\end{eqnarray*}
since $\bar b_n$ is continuous.  We use the notation $A^\eps = \cup_{x\in A} B_\eps(x)$ here.   Thus it remains to prove that for sufficiently large~$n$,
\begin{eqnarray*}
\{x:\bar b_n = 0\}&\subset&\{x: E[b_{\rcs A}(x)]= 0\}^\eps.
\end{eqnarray*}
This follows almost exactly as in \citet[Theorem 1]{cuevas:g:w:r:06}.

By contradiction, suppose that there exists a sequence $x_n \in \{x:\bar b_n = 0\}$ such that $d(\{x_n\}, \{x: E[b_{\rcs A}(x)]= 0\}) >~\eps$ for all $n$.  By compactness, there exists an $x_0$ such that $x_n\rightarrow x_0$, and by continuity, we have that $E[b_{\rcs A}(x_0)]=0$, almost surely.  Therefore, $d(\{x_n\}, \{x: E[b_{\rcs A}(x)]= 0\}) \leq |x_n-x_0|\rightarrow 0,$ which is a contradiction.
\end{proof}

\begin{proof}[Proof of Proposition \ref{prop:understanding_consistency}]
The statements are immediate from definitions and continuity of $E[b_{\rcs A}(x)]$.
\end{proof}

\begin{proof}[Proof of Proposition \ref{prop:equi}]
Let $\bar b_n$ denote the average ODF for the observed sets $A_i,$ and $\bar b_n^1$ denote the average ODF for the observed sets $\alpha A_i.$  For any $\alpha>0$, we have $b_{\alpha A}(x) = \alpha b_A(x/\alpha)$.  It follows that $\bar b_n^1(x)= \alpha \bar b_n(x/\alpha)$ and $E[b_{\rcs A_1}(x)]=\alpha E[b_{\rcs A}(x/\alpha)]$. Next,
\begin{eqnarray*}
\ZZ_n^1(x) &=& \sqrt{n}(\bar b_n^1(x)-E[b_{\rcs A_1}(x)])\\
&=& \alpha \sqrt{n}(\bar b_n-E[b_{\rcs A}])(x/\alpha).
\end{eqnarray*}
Therefore, $\ZZ_n^1(x)\Rightarrow \ZZ^1(x) \stackrel{d}{=} \alpha \ZZ(x/\alpha).$  Lastly, note that
\begin{eqnarray*}
\sup_{x\in \mc W_1}\ZZ^1(x)\stackrel{d}{=} \sup_{x\in \alpha \mc W} \alpha \ZZ(x/\alpha) = \alpha \sup_{x\in \mc W} \ZZ(x).
\end{eqnarray*}
Therefore, a confidence region for $E[\rcs A_1]\cap \mc W_1$ is
\begin{eqnarray*}
&&\hspace{-0.5cm}\{x\in \mc W_1: \bar b_n^1(x)\leq \alpha q_1/\sqrt{n}\} \\
&=& \{x\in \alpha \mc W: \alpha \bar b_n(x/\alpha)\leq \alpha q_1/\sqrt{n}\}\\
&=& \{x\in \alpha \mc W:\bar b_n(x/\alpha)\leq  q_1/\sqrt{n}\}\\
&=& \alpha \{x\in \mc W:\bar b_n(x)\leq  q_1/\sqrt{n}\},
\end{eqnarray*}
and similarly for $\Gamma[\rcs A_1]\cap \mc W_1.$

The same argument works for a rigid motion $g$, since $b_{g(A)}(x) = b_A(g^{-1}(x))$.
\end{proof}

\end{document}